\documentclass[journal]{IEEEtran}
\usepackage{caption}  
\usepackage{cite}
\usepackage{graphicx}
\usepackage[lowtilde]{url}
\usepackage[font=footnotesize]{subfig}
\usepackage{pgfplots}
\pgfplotsset{compat=newest}
\usetikzlibrary{plotmarks}
\usepackage{grffile}
\usepackage{xcolor}
\usepackage{amsmath}  
\usepackage{amsfonts} 
\usepackage{multirow}
\usepackage{enumitem}
\usepackage[euler]{textgreek}
\usepackage{booktabs}
\usepackage{array}
\usepackage{verbatim}
\makeatletter
\long\def\@makecaption#1#2{\ifx\@captype\@IEEEtablestring
\footnotesize\begin{center}{\normalfont\footnotesize #1}\\
{\normalfont\footnotesize\scshape #2}\end{center}
\@IEEEtablecaptionsepspace
\else
\@IEEEfigurecaptionsepspace
\setbox\@tempboxa\hbox{\normalfont\footnotesize {#1.}~~ #2}
\ifdim \wd\@tempboxa >\hsize
\setbox\@tempboxa\hbox{\normalfont\footnotesize {#1.}~~ }
\parbox[t]{\hsize}{\normalfont\footnotesize \noindent\unhbox\@tempboxa#2}
\else
\hbox to\hsize{\normalfont\footnotesize\hfil\box\@tempboxa\hfil}\fi\fi}
\makeatother
\usepackage{amsmath}

\newcommand{\RN}[1]{%
  \textup{\uppercase\expandafter{\romannumeral#1}}%
}

\newcolumntype{C}[1]{>{\centering\let\newline\\\arraybackslash\hspace{0pt}}m{#1}}

\usepackage{hyperref}
\hypersetup{colorlinks=true,
            linkcolor=blue,
            anchorcolor=blue,
            citecolor=blue,
            urlcolor=black}

\begin{document}
\title{Perceptual Quality Assessment of Octree-RAHT Encoded 3D Point Clouds}
\author{Dongshuai~Duan,~Honglei~Su,~\IEEEmembership{Member,~IEEE,}~Qi~Liu,~Hui~Yuan,~\IEEEmembership{Senior~Member,~IEEE,}~Wei~Gao,~\IEEEmembership{Senior~Member,~IEEE,}~Jiarun~Song,~\IEEEmembership{Member,~IEEE}~and~Zhou~Wang,~\IEEEmembership{Fellow,~IEEE}

\thanks{ 
This work was supported in part by the National Science Foundation of China under Grant (62222110, 62172259, 62401307 and 62311530104), in part by the High-end Foreign Experts Recruitment Plan of Chinese Ministry of Science and Technology under Grant G2023150003L, in part by the Taishan Scholar Project of Shandong Province (tsqn202103001), in part by the Shandong Provincial Natural Science Foundation, China, under Grants (ZR2022MF275, ZR2022QF076, ZR2022ZD38 and ZR2021MF025), and in part by Funds for Visiting and Training of Teachers in Ordinary Undergraduate Universities in Shandong Province.

Dongshuai Duan, Honglei Su and Qi Liu are with the College of Electronics and Information, Qingdao University, Qingdao, 266071, China (email: dsduan00@163.com, suhonglei@qdu.edu.cn, sdqi.liu@gmail.com).

Hui Yuan is with the School of Control Science and Engineering, Shandong University, Ji'nan, 250061, China (e-mail: huiyuan@sdu.edu.cn).

Wei Gao is with the School of Electronic and Computer Engineering, Peking University, Shenzhen 518055, China, and also with the Peng Cheng Laboratory, Shenzhen 518066, China (e-mail: gaowei262@pku.edu.cn).

Jiarun Song is with the School of Telecommunications Engineering, Xidian University, Xi’an 710071, China (e-mail: jrsong@xidian.edu.cn).

Zhou Wang is with the Department of Electrical and Computer Engineering, University of Waterloo, Waterloo, ON N2L 3G1, Canada (e-mail: zhou.wang@uwaterloo.ca).

 Corresponding author: Honglei Su.
}}


\markboth{PREPRINT OF A MANUSCRIPT CURRENTLY UNDER REVIEW}
{Shell \MakeLowercase{\textit{et al.}}: Bare Demo of IEEEtran.cls for Journals}

\maketitle

\begin{abstract} 
No-reference bitstream-layer point cloud quality assessment (PCQA) can be deployed without full decoding at any network node to achieve real-time quality monitoring. In this work, we focus on the PCQA problem dedicated to Octree-RAHT encoding mode. First, to address the issue that existing PCQA databases have a small scale and limited distortion levels, we establish the WPC5.0 database which is the first one dedicated to Octree-RAHT encoding mode with a scale of 400 distorted point clouds (PCs) including 4 geometric multiplied by 5 attitude distortion levels. Then, we propose the first PCQA model dedicated to Octree-RAHT encoding mode by parsing PC bitstreams without full decoding. The model introduces texture bitrate (TBPP) to predict texture complexity (TC) and further derives the texture distortion factor. In addition, the Geometric Quantization Parameter (PQS) is used to estimate the geometric distortion factor, which is then integrated into the model along with the texture distortion factor to obtain the proposed PCQA model named streamPCQ-OR. The proposed model has been compared with other advanced PCQA methods on the WPC5.0, BASICS and M-PCCD databases, and experimental results show that our model has excellent performance while having very low computational complexity, providing a reliable choice for time-critical applications. To facilitate subsequent research, the database and source code will be publicly released at \href{https://github.com/qdushl/Waterloo-Point-Cloud-Database-5.0}{https://github.com/qdushl/Waterloo-Point-Cloud-Database-5.0}.
\end{abstract}


\IEEEpeerreviewmaketitle

\section{Introduction}\label{sec:introduction}

\IEEEPARstart {3}{D} point clouds (PCs) are collections composed of a multitude of discrete points in three-dimensional space, each possessing its Cartesian coordinates. Depending on the acquisition method and purpose, each point may also have additional attribute information such as color, normal vectors and reflectance intensity. PCs can be obtained through high-precision 3D scanning technologies like LiDAR and structured light scanning, used to accurately represent the three-dimensional shape and characteristics of objects, scenes, or environments. Given their advantages in providing stereoscopic visual experiences and rich spatial information, point cloud technologies have shown broad application potential in various fields, including virtual reality, augmented reality, game development, autonomous driving and cultural heritage preservation~\cite{gu20193d, 9234526}.

\begin{figure*}[tpb]
\centering
\scalebox{0.9}{
\includegraphics[width=16cm, height=8cm]{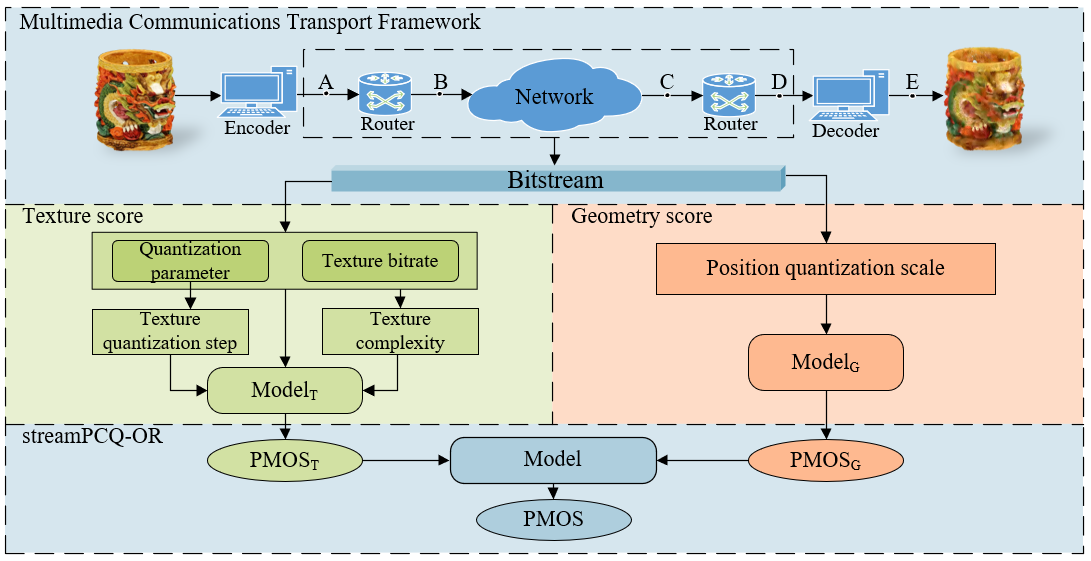}}
\caption{The framework of the proposed streamPCQ-OR model. PMOS is the predicted score of the model.}
\label{fig:Framework}
\end{figure*}

However, PC data processing and transmission face several technical challenges~\cite{9194311, 9106052, liu2019comprehensive}. Firstly, the sheer volume of PC data necessitates efficient compression algorithms to reduce storage and transmission costs. Secondly, the compression process may introduce distortion, negatively impacting the visual and geometric accuracy of PCs. Additionally, PC quality is significantly influenced by the precision of data acquisition devices, environmental factors and post-processing algorithms, complicating the complexity of PCQA. Therefore, to ensure the accuracy and reliability of PC data during compression, transmission and reconstruction, the development and application of reliable PCQA methods are particularly necessary. These methods can help monitor and evaluate potential distortions that may occur during these processes, thereby enhancing the credibility and utility of PC data.

Since humans are the ultimate recipients of various PC applications, subjective quality assessment is undoubtedly the most reliable method for PCQA. However, this method often consumes a considerable amount of time and labor, making it inefficient. Thus, establishing an efficient objective PCQA method is particularly important. Current objective PCQA methods can be categorized into full-reference (FR), reduced-reference (RR) and no-reference (NR) quality assessments, depending on whether the original PC is fully required, partially required, or not required at all. NR methods, with their flexibility of not needing the original PC, have been widely applied. Depending on whether the distorted PC or the compressed bitstream is needed, NR can be further divided into media-based and bitstream-based models. Compared to the former, the latter does not require full decoding, which means significantly reduced processing costs. Fig.~\ref{fig:Framework} illustrates the framework of immersive media communication and the proposed streamPCQ-OR model formulated based on the framework. Media-based NR methods can be deployed at point E after the PC is fully decoded, while bitstream-based NR methods can be deployed at any of points A, B, C, or D giving the latter a higher priority in time and enabling PC quality monitoring. By analyzing the framework of immersive media communication, it can also be understood that there are two main sources of PC distortion: compression distortion and transmission errors. Current PCQA mainly focuses on compression distortion, for the large volume of PC data, compression is an essential process and inevitably introduces compression distortion.

Point cloud compression (PCC) is a research field that has emerged with the rapid development of 3D scanning technology and its applications in recent years. In 2017, the Moving Picture Experts Group (MPEG) initiated the standardization of PCC to meet the growing demand for the storage and transmission of 3D data. In the MPEG-PCC standard, there are two core technologies: Geometry-based Point Cloud Compression (G-PCC)~\cite{ISOIEC2019GPCC} and Video-based Point Cloud Compression (V-PCC)~\cite{ISOIEC2018VPCC}. G-PCC is primarily suitable for static and dynamic acquisition PCs, while V-PCC is designed for dynamically changing PCs. The G-PCC standard defines three geometric encoding modes, including Predictive tree, Octree and Trisoup, as well as three attribute encoding modes, namely Region Adaptive Hierarchical Transform (RAHT), interpolation-based hierarchical nearest-neighbour prediction (Predicting Transform) and 
interpolation-based hierarchical nearest-neighbour prediction with an update/lifting step (Lifting 
Transform)~\cite{whiteG-PCC}. Each encoding mode has its specific applications. Currently, research in the field of PCQA dedicated to Octree-RAHT encoding mode is still in its infancy. To fill this gap, we initiate this in this work.

We first explore the relationship between MOS and distortion parameters, obtaining a preliminary model framework. Then, by fixing the geometric distortion in the geometric and textural distortion of Octree-RAHT encoded PCs and considering the masking effect of the HVS, we predict TC through TBPP and estimate the texture distortion factor. After determining the texture distortion factor, we further estimate the geometric distortion factor through geometric quantization parameters. Finally, by substituting the texture and geometric distortion factors into the model, we obtain the streamPCQ-OR model, which is verified for its predictive performance in the results and analysis. 

The main contributions of this paper can be summarized as follows:

1) We establish the first PCQA database dedicated to Octree-RAHT encoding mode. By selecting a batch of high-quality source PCs and applying the Octree-RAHT encoding mode, 400 distortion PCs containing 4 geometric multiplied by 5 attitude distortion levels were generated. Subjective testing was conducted to obtain Mean Opinion Scores (MOSs). The proposed database was named WPC5.0.

2) We propose the first PCQA model specially designed for Octree-RAHT encoding mode by analyzing PC bitstreams without full decoding. The model introduces texture bitrate (TBPP) to predict texture complexity (TC) and further derives the texture distortion factor. In addition, the Geometric Quantization Parameter (PQS) is used to estimate the geometric distortion factor, which is then
integrated into the model along with the texture distortion factor to obtain the proposed PCQA model named streamPCQ-OR.

3) We validate streamPCQ-OR on the WPC5.0, BASICS~\cite{ak2024basics} and M-PCCD~\cite{alexiou2019comprehensive} database and compared it with existing competitive PCQA models to verify its effectiveness and robustness.

The remainder of this paper is organized as follows. In Section~\ref{2}, we discuss the related work on objective PCQA models from the perspectives of FR, RR and NR. In Section~\ref{3}, we introduce the WPC5.0 database constructed to address the issues of limited training sets and insufficient distortion levels in existing databases, as well as the subsequent subjective experiments conducted to obtain the corresponding MOS for the database. In Section~\ref{4}, we establish the streamPCQ-OR model. In Section~\ref{5}, we report our experimental results and model performance. Finally, we draw conclusions in Section~\ref{6}.

\section{Related Work}\label{2}
Based on the dependency on the original PC, PCQA methods can be categorized into FR, RR and NR. Furthermore, based on whether the reference information comes from the PC itself or the compressed bitstream, PCQA models can be further divided into media-layer models and bitstream-layer models. 
\subsection{FR Models}\label{2A}
 For point-based metrics, the most classic ones are point-to-point measurement (p2po)~\cite{mekuria2016evaluation} and point-to-plane measurement (p2pl)~\cite{tian2017geometric}. P2po evaluates geometric distortion by calculating the Euclidean distance between corresponding points in the original and distorted PCs. P2pl, on the other hand, evaluates geometric distortion by calculating the projection error along the normal vector direction between corresponding points in the original and distorted PCs. Later, $\operatorname{PSNR_Y}$~\cite{mekuria2017performance} was further proposed to assess the texture distortion of PCs. Alexiou \textit{et al}.~\cite{alexiou2020towards} introduced PointSSIM, an assessment method based on statistical dispersion that focuses on capturing local changes to predict visual quality. Liu \textit{et al}.~\cite{liu2022perceptual} proposed $\operatorname{IW-SSIM_P}$, a PCQA model based on the principle of information content-weighted structural similarity. GraphSIM, proposed by Yang \textit{et al}.~\cite{yang2020inferring}, brought innovation to PCQA by simulating human perception of geometry and color impairments. Zhang \textit{et al}.~\cite{zhang2021ms} further proposed MS-GraphSIM, a multi-scale assessment model that improves prediction accuracy through assessments at different scales. Meynet \textit{et al}.~\cite{meynet2020pcqm, meynet2019pc} contributed two measurement methods, PCQM and PC-MSDM, the former considering both geometric and color features, and the latter focusing on local curvature statistics. In addition, Viola \textit{et al}.~\cite{viola2020color} proposed an objective measurement method based on color for the color distortion problem of PCs in immersive scenarios. Hua \textit{et al}.~\cite{hua2020vqa, hua2022cpc} proposed two methods, VQA-CPC and CPC-GSCT, to evaluate the visual quality of color PCs from geometric and texture perspectives. Wang \textit{et al}.~\cite{wang2024integrated} proposes a novel method that incorporates heterogeneous graph attention network to take advantage of information of multiple modalities and significantly improve the representation quality. Differently, CSSR~\cite{huang2022class} designs a novel framework with class-specific auto-encoders to enhance the representation capability of deep neural networks to achieve compact and accurate representations. Javaheri \textit{et al}.~\cite{javaheri2020improving, javaheri2020generalized} focused on improving the reliability of measurements, proposing improved objective quality measurement methods based on PSNR, and evaluated the compression performance of PC coding solutions. Diniz \textit{et al}.~\cite{diniz2020multi, diniz2020towards, diniz2021novel, diniz2020local} made a series of contributions to PC quality measurement, including methods based on distances between multiple PCs, texture-based assessment methods, perceptual color distance patterns (PCDP), and assessment methods using local luminance patterns (LLP). Diniz \textit{et al}.~\cite{diniz2021color} also developed BitDance, a low-complexity full-reference visual quality measurement method suitable for static PC. Xu \textit{et al}.~\cite{xu2021epes} proposed an assessment model based on elastic potential energy similarity (EPES), quantifying the impact of distortion on PC quality by comparing elastic potential energy. 

For projection-based methods, Torlig \textit{et al}.~\cite{torlig2018novel} developed a real-time voxelization and projection framework, using two-dimensional quality measurement methods to predict perceptual quality. Wu \textit{et al}.~\cite{wu2021subjective} studied the impact of geometric and texture attributes in compression distortion and proposed projection-based objective quality assessment methods. He \textit{et al}.~\cite{he2021towards} combined color texture and curvature projection, extracting texture and geometric statistical features to assess the quality of color PCs. Yang \textit{et al}.~\cite{yang2020predicting} predicted PC quality by aggregating global and local image-based features. He \textit{et al}.~\cite{he2022tgp}'s TGP-PCQA method uses 4D tensor decomposition technology and edge feature calculation to assess the distortion of color PC.

For methods based on both points and projections, Tu \textit{et al}.~\cite{tu2023pseudo} proposed a pseudo-reference PCQA metric based on joint 2D and 3D distortion description, providing a new perspective for assessing distortions introduced by PC compression.

\subsection{RR Models}\label{2B}
Viola \textit{et al}.~\cite{viola2020reduced} proposed a RR metric that only needs to extract a small set of statistical features from the reference PC, such as features in the geometry, color and normal vector domains, to assess the visual degradation of the content at the receiving end. Liu \textit{et al}.~\cite{liu2021reduced} proposed a new linear perception quality model, which is fitted based on geometric and color quantization steps, for assessing PCs encoded by V-PCC. Liu \textit{et al}.~\cite{liu2022reduced} developed a RR model called R-PCQA, focusing on quantifying the distortions introduced by lossy compression, and developed the model by analyzing the compression distortion of PCs under separate attribute compression and geometric compression. Liu \textit{et al}.~\cite{10375131} proposed a metric called PCQAML, which employs the Least Absolute Shrinkage and Selection Operator (LASSO) method to select the most effective features from the proposed set of features and further map them to the Mean Opinion Score. Zhou \textit{et al}.~\cite{zhou2023reduced} proposed a RR PC quality metric based on content-oriented saliency projection (RR-CAP), making the first attempt to simplify the reference and distorted PCs into projected saliency maps through downsampling operations, solving the problem of transmitting large volumes of original PC data. Zhang \textit{et al}.~\cite{zhang2024reduced} designed the first RR quality assessment metric specifically for textured mesh 3D digital humans (DHs), calculating geometric curvature-related attributes and texture-related indicators, and using a support vector regression (SVR) model for statistical analysis to achieve quality prediction.

\subsection{NR Models}\label{2C}
For point-based models, Zhang \textit{et al}.~\cite{zhang2022no} proposed a NR visual quality assessment method for 3D models, which projects from 3D space into geometric and color feature domains, and uses 3D natural scene statistics (3D-NSS) and entropy to extract quality-aware features, then regresses these features into visual quality scores through a support vector regression (SVR) model. Liu \textit{et al}.~\cite{liu2023point} proposed a NR metric called ResSCNN based on a sparse convolutional neural network (CNN) to accurately estimate the subjective quality of PCs. Liu \textit{et al}.~\cite{liu2022progressive} proposed PKT-PCQA based on the human visual perception mechanism, which merges local and global features and considers the characteristics of the HVS. Zhou \textit{et al}.~\cite{zhou2024blind} proposed SGR that automatically evaluates the visual quality of dense 3D PCs through structure-guided resampling and feature extraction [43]. Zhu \textit{et al}.~\cite{zhu20243dta} proposed a new two-stage sampling method and a dual-attention-based transformer model (3DTA) that can efficiently and accurately assess PC quality. Shan \textit{et al}.~\cite{Shan_2024_CVPR} proposed a novel contrastive pre-training framework (CoPA), which creates anchors by projecting point clouds with different distortions into images and mixing local areas. The method utilizes a quality-aware contrastive loss for pre-training and introduces a semantic-guided multi-view fusion module to integrate features from multi-perspective projected images.

For projection-based models, Tu \textit{et al}.~\cite{tu2022v} proposed a new blind PCQA metric that uses a dual-stream convolutional network to extract texture and geometry features, and combines visual perception to consider distortions in salient regions. Shan \textit{et al}.~\cite{shan2023gpa} proposed the GPA-Net, introducing a new graph convolution kernel called GPAConv, which focuses on capturing structural and textural perturbations, and uses a multi-task framework for quality regression and prediction of distortion type and degree. Liu \textit{et al}.~\cite{liu2021pqa} proposed PQA-Net, a deep learning-based NR method, composed of a multi-view joint feature extraction and fusion (MVFEF) module, a distortion type identification (DTI) module, and a quality vector prediction (QVP) module. Zhang \textit{et al}.~\cite{zhang2024gms} proposed a projection-based grid mini-patch sampling 3D model quality assessment method (GMS-3DQA), using the Swin-Transformer tiny backbone network to extract quality-aware features from quality mini-patch maps (QMM). Liu \textit{et al}.~\cite{liu2023once} proposed D$^{3}$-PCQA, enhancing model interpretability through a kernel ridge regression model and introducing a description domain to reduce domain discrepancy and improve generalization. Tao \textit{et al}.~\cite{tao2021point} proposed PM-BVQA, a blind visual quality assessment method for color PCs, by projecting PCs onto 2D maps and using a multi-scale feature fusion network to assess quality. Yang \textit{et al}.~\cite{yang2022no} proposed IT-PCQA, using DNN to emulate the evaluation criteria of the HVS. Fan \textit{et al}.~\cite{fan2022no} proposed a NR method based on video sequences of colored PCs, obtaining specific trajectory video sequences by rotating the camera around point clouds, and using a modified ResNet3D model as a feature extraction module. Zhang \textit{et al}.~\cite{zhang2022treating} combines video quality assessment techniques, effectively predicting point cloud visual quality by capturing point cloud videos through dynamic rotation and extracting spatial and temporal features. Wang \textit{et al}.~\cite{wang2024zoom} proposed MOD-PCQA that enhances quality perception through multi-viewpoint projection and multi-scale feature extraction.

Additionally, for methods based on both points and projections, Hua \textit{et al}.~\cite{hua2021bqe} proposed BQE-CVP for color PCs based on visual perception, which considers the impact of visual masking effects on quality assessment and extracts geometric and color combination feature vectors to predict the quality of CPC. Zhang \textit{et al}.~\cite{zhang2022mm} proposed MM-PCQA that integrates 2D and 3D information, by segmenting PCs, extracting texture features, encoding sub-models and projected images, and using cross-modal attention mechanisms to fuse information. Chen \textit{et al}.~\cite{chen2024dynamic} proposed an evaluation method based on hypergraphs. This method uses a dynamic hypergraph convolutional network (DHCN) for feature extraction and quality reasoning, improving the accuracy of quality prediction. Chai \textit{et al}.~\cite{chai2024plain} proposed the Plain-PCQA model, which includes a NR branch and a degraded-reference (DR) branch, as well as a plane-point interaction transformer (P2IT), to assess the visual quality of 3D PCs by combining 2D projections and 3D point descriptors, with the projections processed by a lightweight network ResNet-18 and introducing a mask weight to enhance the accuracy of feature extraction. Mu \textit{et al}.~\cite{MU2024122953} proposed Hallucinated-PQA, a method that employs a distortion recovery network to preprocess images, providing pseudo-reference information for assessment, and designed a Hallucination Injection Block (HIB) and a Multi-View and Multi-Scale Context Fusion (MMCF) module to capture the contextual correlations between multi-view features.

In addition to media-layer PCQA models aforementioned, there are also bitstream-layer models. Liu \textit{et al}.~\cite{liu2022no} proposed a NR bitstream-layer model for assessing the perceptual quality of V-PCC encoded PCs, establishing texture and geometry models by analyzing texture and geometric parameters respectively and further combining them into an overall model. Su \textit{et al}.~\cite{su2023bitstream} developed a bitstream-based NR model for assessing the perceptual quality of compressed PCs, without full decoding, by establishing a relationship between texture complexity and bitrate and texture quantization parameters to assess texture distortion. Lv \textit{et al}.~\cite{10637704} proposed a bitstream-based NR model, which predicts the visual quality of PCs by analyzing texture quantization parameters, texture complexity, and position quantization scales. However, there has been no research discussing the PCQA problem dedicated to Octree-RAHT encoding mode yet.
\subsection{Summery}\label{2D}
In the final analysis, whether a FR, RR or NR method, it can effectively assess the PC quality. However, given the difficulty in obtaining reference PCs in many practical applications, current researches in PCQA filed primarily focus on developing NR methods. It is not difficult to observe that most existing NR-PCQA models belong to the media-layer, which unlike a bitstream-layer model, cannot be flexibly embedded into immersive media communication. In addition, bitstream-layer models have an inherent advantage of low computational complexity due to their partial decoding nature, making them more suitable for real-time applications. We focus on the PCQA issue dedicated to Octree-RAHT encoding mode in this work. For detailed description, please refer to Section~\ref{3} and~\ref{4}.

\section{The WPC5.0 Database}
\label{3}
 Existing databases~\cite{alexiou2019comprehensive, alexiou2020pointxr, ak2024basics} including Octree-RAHT encoded PCs are either small-scale or have limited distortion levels, which can easily lead to overfitting or excessive bias during the model training. To address this issue, we established the first PCQA database dedicated to Octree-RAHT encoding mode in this Section. 
\subsection{Database Establishment}
\subsubsection{Original contents}
We selected Bag, Banana, Biscuits, Cake, Cauliflower, Flowerpot, Glasses\_case, Honeydew\_melon, House, Litchi, Mushroom, Pen\_container, Pineapple, Ping-pong\_bat, Puer\_tea, Pumpkin, Ship, Statue, Stone and Tool\_box from the Waterloo Point Cloud database (WPC)~\cite{liu2022perceptual}. With an average point size of 1.35M and a standard deviation of 656K, these 20 3D source PCs are rich in geometric and textural details, making them quality samples for PC processing and evaluation.

\subsubsection{Distorted contents}
The implementation of G-PCC we use is the TMC13 V20.0, by selecting Octree-RAHT as the encoding mode. Geometric positional quantisation scales are selected as 0.125, 0.25, 0.5 and 1, and the attribute quantisation parameters are selected as 22, 28, 34, 40 and 46. The rest of the encoding parameters are set as default values. For each source PC in WPC5.0, \(4\times5\) = 20 distortion levels can be generated, resulting in a total of \(20\times4\times5\) = 400 distorted PCs.

\subsection{Subjective Test}
\subsubsection{Experimental settings}
\begin{figure}\centering
{\includegraphics[width=0.8\linewidth]{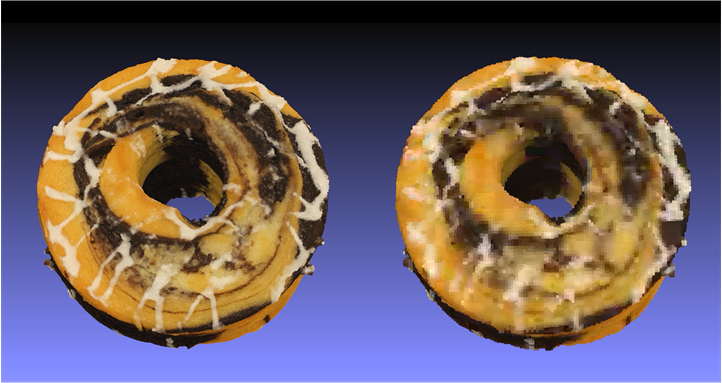}}\hskip0.1em
\caption{ Display of original and distorted PCs.}
\label{fig:screen}
\end{figure}

Firstly, the WPC5.0 database and 20 original PCs were rendered using Pccrenderer (version V.5). The parameter settings used for rendering are shown in the Table~\ref{tab:parameter settings}. A total of 400 videos were rendered to generate 400 videos, in which the geometric centre of the PC was used as the centre of the circle for the rendering, and the horizontal and vertical circles were selected as the virtual camera paths to continuously generate the point of view on these circles, and a video sequence was generated for each distorted PC and original PC horizontally connected. 
\begin{table}[t]
\centering
\caption{Rendering Parameter Settings}
\label{tab:parameter settings}
\scalebox{1.0}{
\begin{tabular}{c | c }
\hline
\hline
Setting items & Parameter\\
\hline
Render window & \(960\times960\) \\
Point type & 'point' \\
Point size ($\operatorname{PQS}_{0.125, 0.25, 0.5, 1}$) & 8, 4, 2, 1 \\
Virtual camera radius & 5000 \\
Degree/Point of view & 2 \\
Frame image & \(180 fps\times2\) \\
Video sequence time & 10 seconds \\
Screen resolution & \(1920\times1080\) \\
Screen Size & 23.8 inches \\
Screen refresh rate & 60HZ \\
\hline
\hline
\end{tabular}
}
\end{table}
Each connection between the distorted PC and the original PC level generates a video sequence for display, and all video sequences are shown using the display with the parameters in the Table~\ref{tab:parameter settings}. The video part of the interface is shown in the Fig.~\ref{fig:screen}, with the original reference PC in the left half and the distorted PC in the right half.
The subjective tests were performed based on the BT-500.13 standard~\cite{series2012methodology}. We used the subjective evaluation procedure for the test, where subjects can watch and rate the videos, all of which were displayed in native resolution to avoid further scaling distortion. 60 Hz refresh rate was greater than the frame rate of all videos. Prior to the study, we carefully tested the playback of all PC rendering sequences to eliminate any issues related to latency, dropped frames, loss of synchronisation, etc.

A total of 30 students of different ages and genders from Qingdao University participated in this test, and all testers had normal vision or corrected-to-normal vision. The subjective evaluation was conducted using the double-stimulus impairment scale (DSIS) method~\cite{YanaComparison}. The video frame is shown in the Fig.~\ref{fig:screen}, and the subjects score by observing the standard PC in the left half of the video and the distorted PC in the right half, with a rating range of 0-100. In order for the testers to understand the various distortions of the PC before the formal test, a training session was set up and the testers were shown 20 distorted PCs with different PCs and different encoding parameters. Each participant will score 400 distorted PCs to produce 400 scoring data after a subjective test of an average length of 2 hours. To prevent visual fatigue caused by the participant looking at the screen for a long period of time, the whole test was divided into 4 sessions, and a 5-minute break was set up after every 100 PCs were shown. In the end, each PC corresponded to 30 subjective scores, so a total of 12,000 subjective scores were obtained.

\subsubsection{Experimental results}
\begin{figure}[tpb]
\centering
\subfloat[]
{\includegraphics[width=0.49\linewidth]{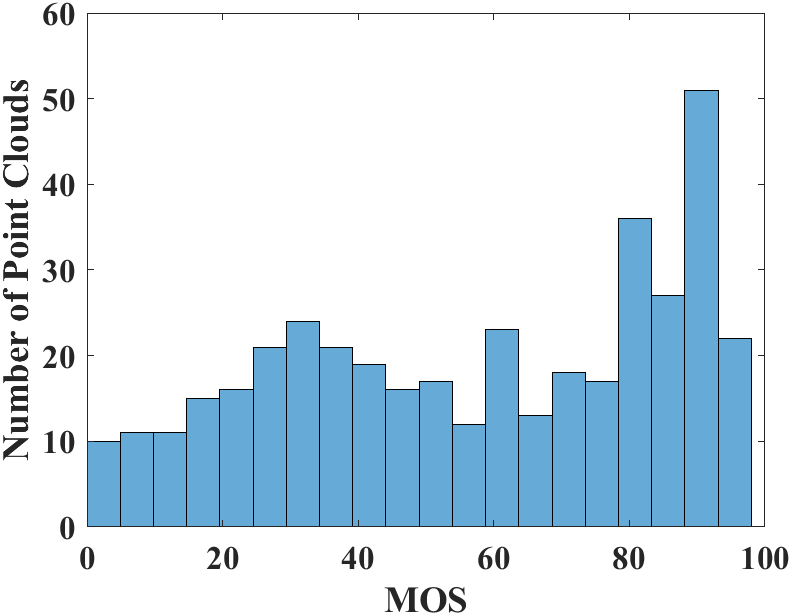}\label{fig:3a}}\hskip0.1em
\subfloat[]
{\includegraphics[width=0.49\linewidth]{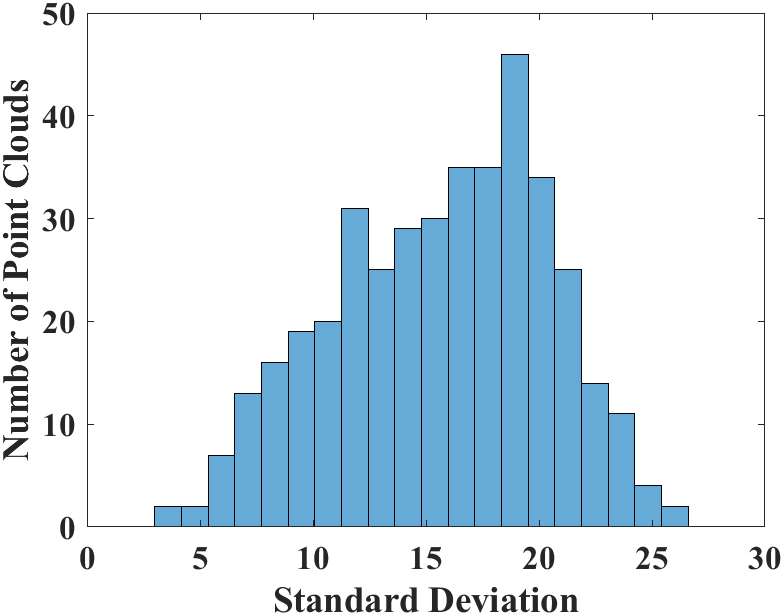}\label{fig:3b}}\hskip0.1em
\caption{MOS statistics of the WPC5.0 dataset.}
\label{fig:MOS distribution}
\end{figure}

\begin{figure}[tpb]
\centering
\subfloat[]
{\includegraphics[width=0.49\linewidth]{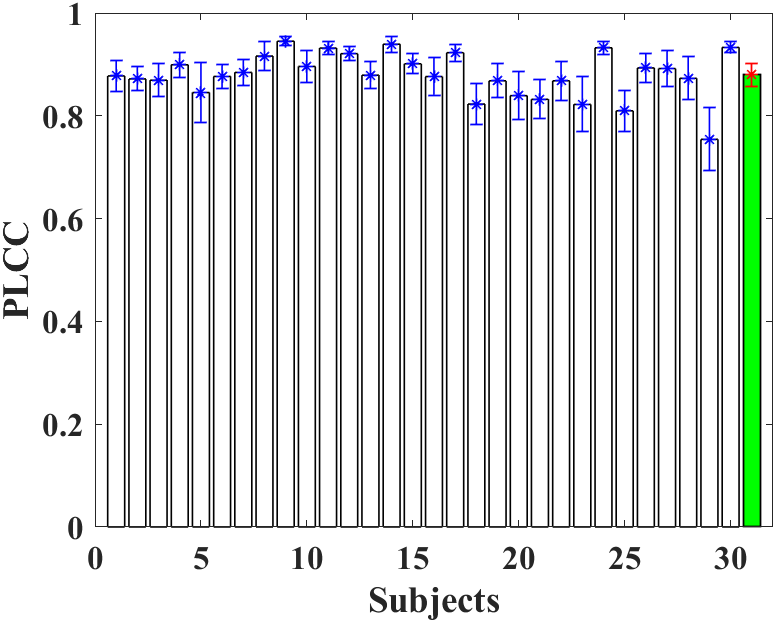}\label{fig:4a}}\hskip0.1em
\subfloat[]
{\includegraphics[width=0.49\linewidth]{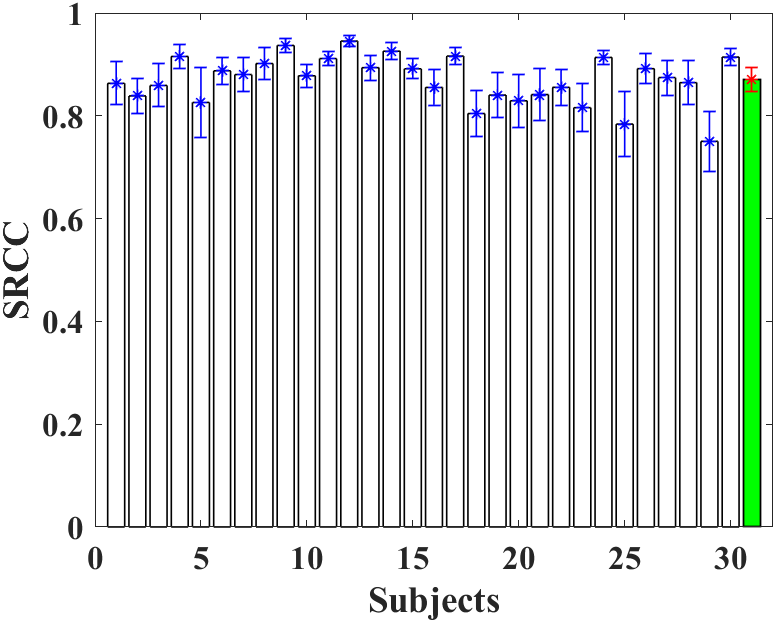}\label{fig:4b}}\hskip0.1em
\caption{PLCC and SRCC between individual subject ratings and MOS. Rightmost column: performance of an average subject.}
\label{fig:MOS_srccplcc}
\end{figure}

In order to explore the reasonableness of the data obtained from the subjective test, we conducted further data analysis and discussion. After the test we converted the subjective scores into Z-scores as suggested in the literature~\cite{van1995quality} and used an outlier removal scheme~\cite{series2012methodology} to detect the presence of outliers. For the m-th PC at the j-th level of degradation as perceived by the i-th observer, the Z-score is calculated as \begin{equation}
Z_{m,i,j} = \frac{X_{m,i,j} - \mu_{Xi}}{\delta_{Xi}},
\label{equ:ZMOS}
\end{equation}in which $X_{m,i,j}$ denotes the original score, while $\mu_{Xi}$ and $\delta_{Xi}$ correspond to the average and the standard deviation of the i-th observer's ratings, respectively. The Z-scores linear scaling was then adjusted to the range [0,100]. The MOS of each distorted PC was calculated by averaging all valid and scaled Z-scores.The histogram of the MOS distribution is shown in Fig.~\ref{fig:3a}. It is clear that the distorted PC quality scores are distributed over a wide range. The standard deviation of the scores of all testers for each distorted PC is shown in Fig.~\ref{fig:3b}. It is clear that the standard deviation is small, indicating that the overall testers performed well. We calculate the Pearson Linear Correlation Coefficient (PLCC) and Spearman Rank-order Correlation Coefficient (SRCC) between each tester's score and MOS to estimate the performance of individual subjects, and the results are presented in Fig.~\ref{fig:MOS_srccplcc}. The average performance of all individual subjects is also given in the rightmost column of the figure. The mean values of PLCC and SRCC between each subject and MOS are as high as 0.8797 and 0.8701, respectively, indicating that most of the individual performances are basically consistent with each other.

\section{Proposed streamPCQ-OR Model}
\label{4}
\subsection{Exploring the Relationship Between MOS and Distortion Parameters}
\label{4A}
Due to the fact that Octree-RAHT distorted PCs encompass geometric distortions resulting from Octree encoding and textural distortions stemming from RAHT encoding, for ease of analysis, we initially fix the geometric distortion to study the relationship between MOS and textural coding parameters under various geometric distortion conditions. Similar to image and video compression, the encoding distortion caused by RAHT in reconstructed PCs is primarily manifested as blocking artifacts. This is caused by the independent quantization of RAHT coefficients across adjacent virtual surfaces. Although the post-processing algorithm of G-PCC has undergone quality enhancement, it still appears blurry. Therefore, texture quantization is the main cause of textural distortion, and perceived textural distortion is closely related to the texture quantization step size or the corresponding quantization parameter (QP). 
\begin{figure}[tpb]
\centering
\subfloat[PQS=1]
{\includegraphics[width=0.49\linewidth]{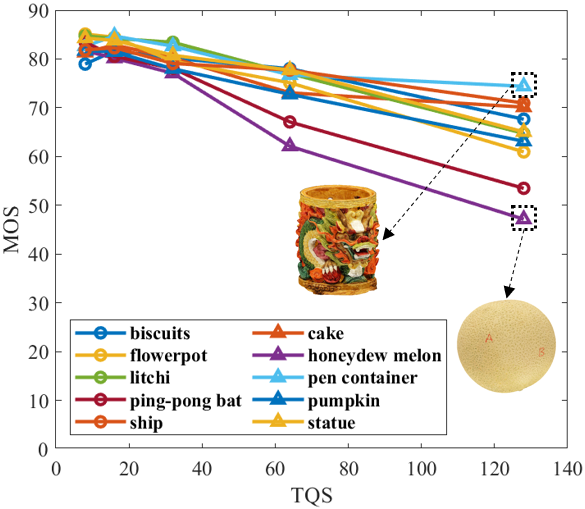}}\hskip0.1em
\subfloat[PQS=0.5]{\includegraphics[width=0.49\linewidth]{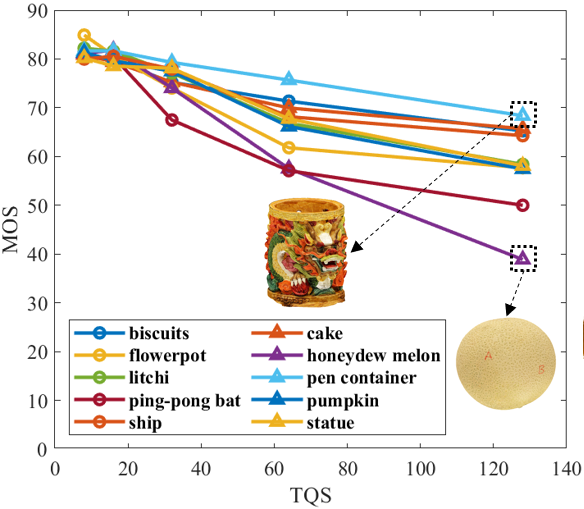}}\hskip0.1em \subfloat[PQS=0.25]{\includegraphics[width=0.49\linewidth]{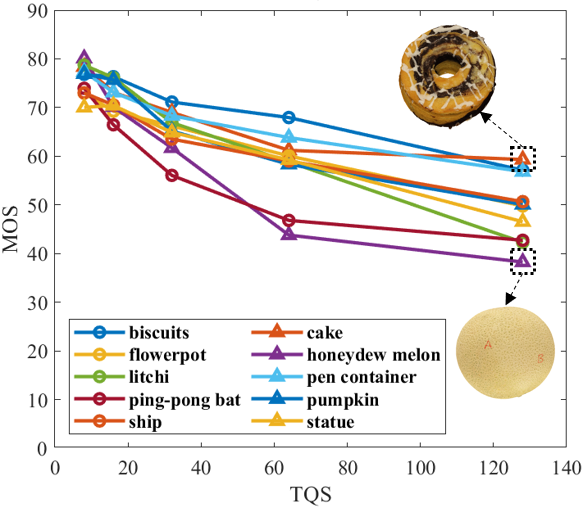}}\hskip0.1em
\subfloat[PQS=0.125]{\includegraphics[width=0.49\linewidth]{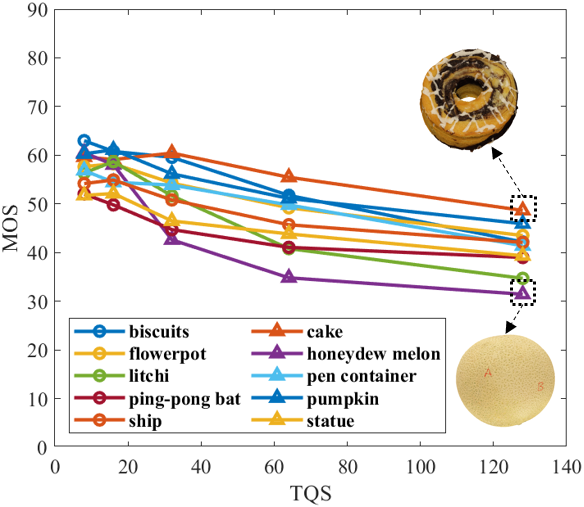}}\hskip0.1em
\caption{Relationship between MOS and TQS at different PQS.}
\label{fig:MOS_TQS}
\end{figure}
To explore the underlying relationship, we randomly selected ten categories of PCs from the WPC5.0 database, calculated the PLCC between MOS and texture quantization step (TQS) under fixed geometric distortion and varying textural distortion conditions, and plotted line charts for visual observation. \begin{table}[t]  
\centering
\caption{PLCC between MOS and TQS in Different PC. $\operatorname{PQS}_{1/0.5/0.25/0.125}$ denotes different geometric distortions.}
\label{tab:MOSTQSplcc}
\scalebox{0.9}{
\begin{tabular}{c | c c c c}  
\hline
\hline
\multirow{2}{*}{Name}&
  \multicolumn{4}{c}{PLCC}\\
  \cline{2-5} 
&$\operatorname{PQS}_{1}$&$\operatorname{PQS}_{0.5}$& $\operatorname{PQS}_{0.25}$ &$\operatorname{PQS}_{0.125}$\\
\hline  
\emph{Biscuits}        &0.9330&0.9770&0.9930&0.9950\\
\emph{Cake}            &0.9440&0.9620&0.8960&0.9660\\
\emph{Flowerpot}       &0.9970&0.9280&0.9930&0.9820\\
\emph{Honeydew\_melon} &0.9910&0.9910&0.9160&0.8700\\
\emph{Litchi}          &0.9910&0.9810&0.9900&0.9530\\
\emph{Pen\_container}  &0.9300&0.9980&0.9630&0.9950\\
\emph{Ping-pong\_bat}  &0.9960&0.9350&0.8900&0.9000\\
\emph{Pumpkin}         &0.9980&0.9830&0.9520&0.9710\\
\emph{Ship}            &0.9870&0.9510&0.9690&0.9600\\
\emph{Statue}          &0.9920&0.9870&0.9960&0.9520\\
\hline
Mean                   &0.9932&0.9915&0.9854&0.9726\\
\hline
\hline
\end{tabular}
}
\end{table}
Fig.~\ref{fig:MOS_TQS} illustrates that as the TQS increases, the MOS gradually decreases, indicating that the distortion of the PC increases with the enlargement of the texture quantization step, leading to a decline in quality scores. Simultaneously, it can be observed that the slopes of the MOS versus TQS curves vary among different PCs, this phenomenon is attributed to the important role played by the texture masking effect of the HVS in texture distortion assessment. The texture blurring caused by RAHT becomes increasingly apparent as the TQS increases. However, PCs with rich texture details can mitigate the impact of such texture blurring. Consequently, PCs with higher texture complexity are likely to achieve higher quality scores at the same TQS, such as "Pen\_container" and "Cake". Whereas those with lower texture complexity tend to receive lower quality scores, such as "Honeydew\_melon."

Moreover, it is essential to recognize that under scenarios of substantial geometric distortion, the influence of geometric distortion predominates, thereby reducing the effect of variations in texture detail on quality scores. The differences in the slopes of the MOS versus TQS curves among various PCs become inconspicuous, with the exception of those displaying marked disparities in texture detail, such as "Cake" and "Honeydew\_melon", where the slopes of the MOS versus TQS curves remain distinctly divergent. For PCs with negligible differences in texture detail, the slope variations of the MOS versus TQS curves fade, and the curve intercepts start to exhibit discrepancies. In the case of PCs with minor variations in texture detail, as geometric distortion escalates, those with simpler geometries are more prone to securing higher scores. This is attributable to the fact that simpler geometries exhibit reduced sensitivity to distortions caused by Octree encoding. Consequently, under conditions of elevated geometric distortion, a geometrically straightforward PC such as "Cake" achieves a higher score compared to the geometrically intricate "Pen\_container."
\begin{figure*}
    \centering
\captionsetup{justification=centering}
    \subfloat[PQS=1 QP=22]{\includegraphics[width=0.19\linewidth]{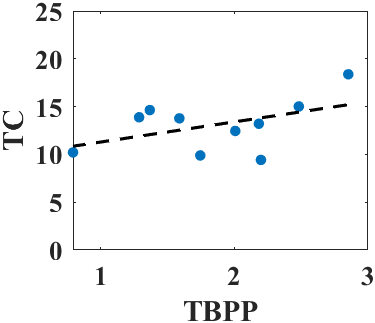}}\hskip0.1em
    \subfloat[PQS=1 QP=28]{\includegraphics[width=0.19\linewidth]{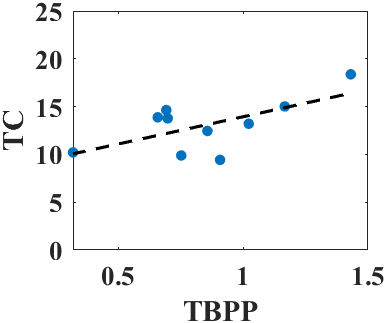}}\hskip0.1em
    \subfloat[PQS=1 QP=34]{\includegraphics[width=0.19\linewidth]{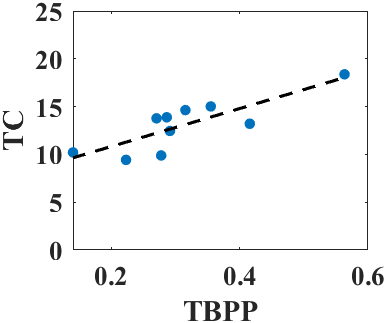}}\hskip0.1em
    \subfloat[PQS=1 QP=40]{\includegraphics[width=0.19\linewidth]{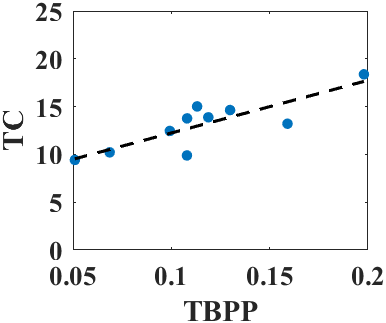}}\hskip0.1em 
    \subfloat[PQS=1 QP=46]{\includegraphics[width=0.19\linewidth]{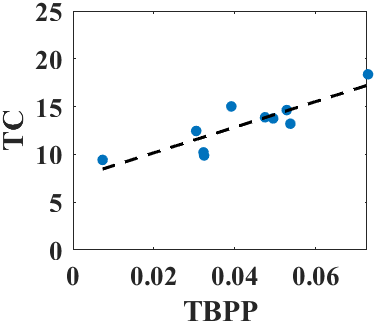}}\hskip0.1em
    \subfloat[PQS=0.5 QP=22]{\includegraphics[width=0.19\linewidth]{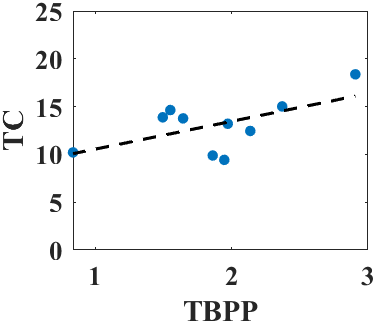}}\hskip0.1em
    \subfloat[PQS=0.5 QP=28]{\includegraphics[width=0.19\linewidth]{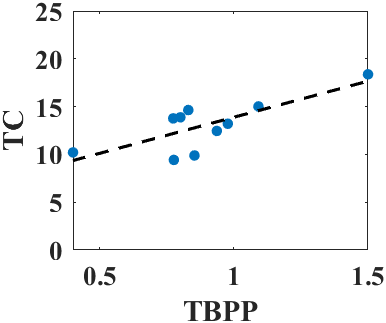}}\hskip0.1em
    \subfloat[PQS=0.5 QP=34]{\includegraphics[width=0.19\linewidth]{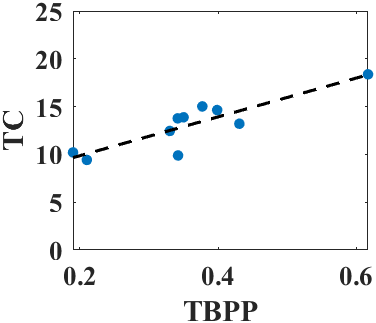}}\hskip0.1em
    \subfloat[PQS=0.5 QP=40]{\includegraphics[width=0.19\linewidth]{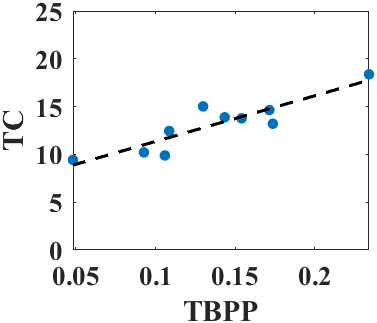}}\hskip0.1em
    \subfloat[PQS=0.5 QP=46]{\includegraphics[width=0.19\linewidth]{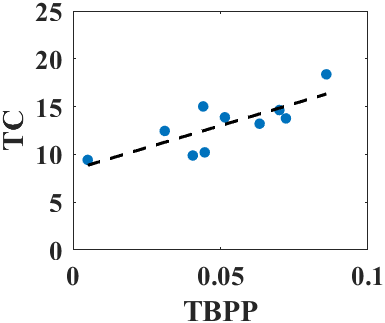}}\hskip0.1em
    \subfloat[PQS=0.25 QP=22]{\includegraphics[width=0.19\linewidth]{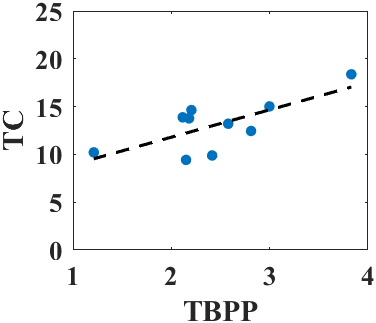}}\hskip0.1em
    \subfloat[PQS=0.25 QP=28]{\includegraphics[width=0.19\linewidth]{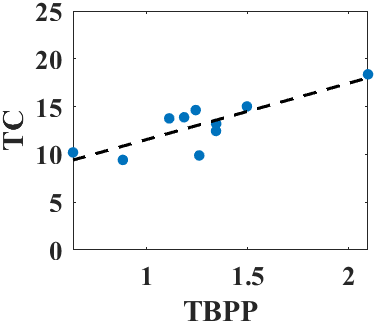}}\hskip0.1em
    \subfloat[PQS=0.25 QP=34]{\includegraphics[width=0.19\linewidth]{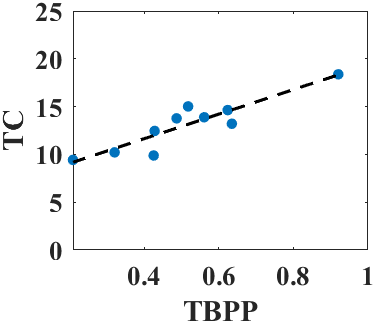}}\hskip0.1em
    \subfloat[PQS=0.25 QP=40]{\includegraphics[width=0.19\linewidth]{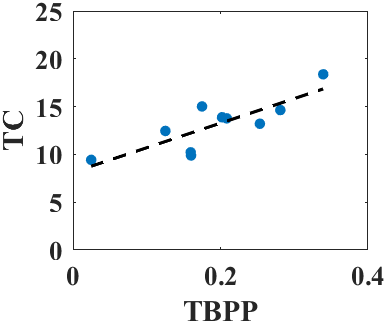}}\hskip0.1em
    \subfloat[PQS=0.25 QP=46]{\includegraphics[width=0.19\linewidth]{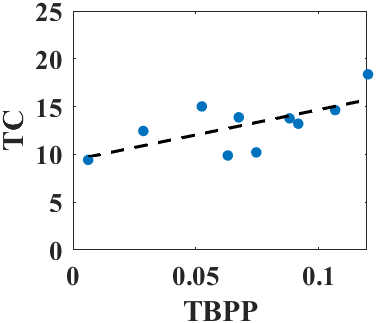}}\hskip0.1em
    \subfloat[PQS=0.125 QP=22]{\includegraphics[width=0.19\linewidth]{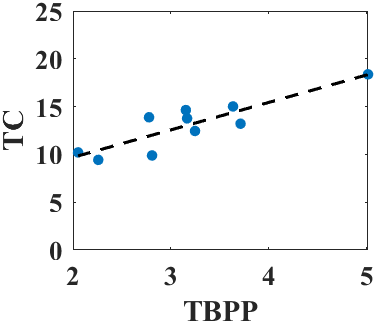}}\hskip0.1em
    \subfloat[PQS=0.125 QP=28]{\includegraphics[width=0.19\linewidth]{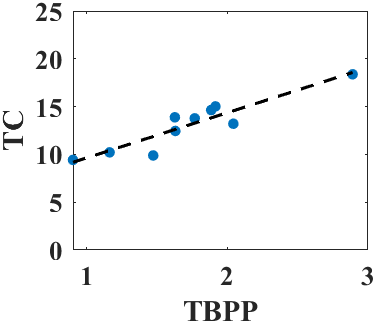}}\hskip0.1em
    \subfloat[PQS=0.125 QP=34]{\includegraphics[width=0.19\linewidth]{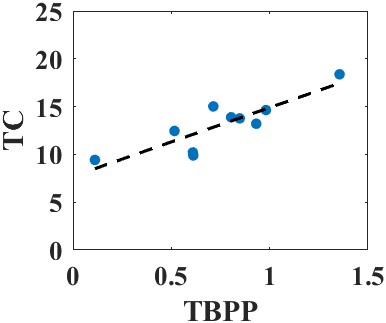}}\hskip0.1em
    \subfloat[PQS=0.125 QP=40]{\includegraphics[width=0.19\linewidth]{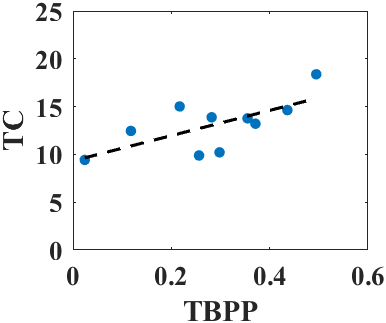}}\hskip0.1em
    \subfloat[PQS=0.125 QP=46]{\includegraphics[width=0.19\linewidth]{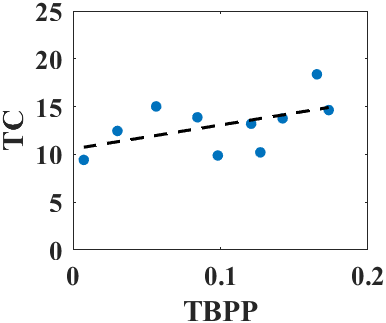}}\hskip0.1em
    \caption{Relationship between TBPP and TC at different PQS and QP.}
    \label{fig:TBPP vs. TC}
\end{figure*}
\begin{figure*}
    \centering
    \subfloat[PQS=1]{\includegraphics[width=0.20\linewidth]{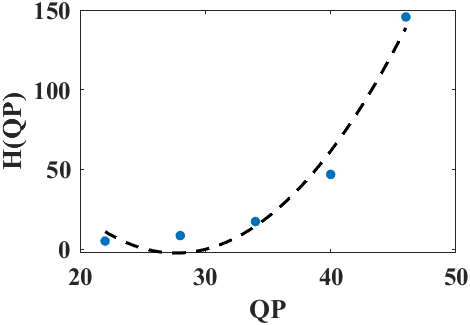}}\hspace{3mm}
    \subfloat[PQS=0.5]{\includegraphics[width=0.20\linewidth]{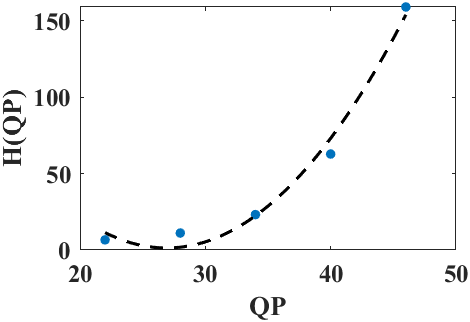}}\hspace{3mm}
    \subfloat[PQS=0.25]{\includegraphics[width=0.2\linewidth]{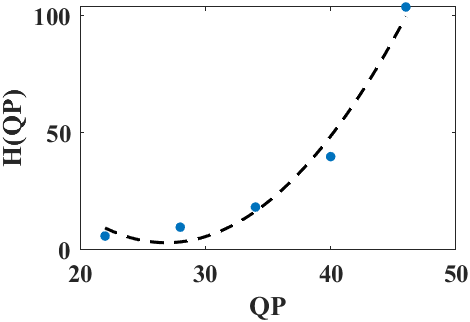}}\hspace{3mm}
    \subfloat[PQS=0.125]{\includegraphics[width=0.2\linewidth]{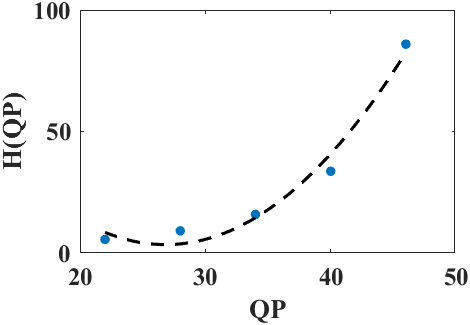}}\hspace{3mm}
    \caption{Relationship between H(QP) and QP at different PQS.}
    \label{fig:H vs TQP}
\end{figure*}
\begin{figure*}
    \centering
    \subfloat[PQS=1]{\includegraphics[width=0.2\linewidth]{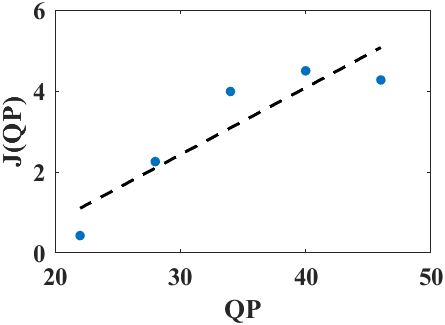}}\hspace{3mm}
    \subfloat[PQS=0.5]{\includegraphics[width=0.2\linewidth]{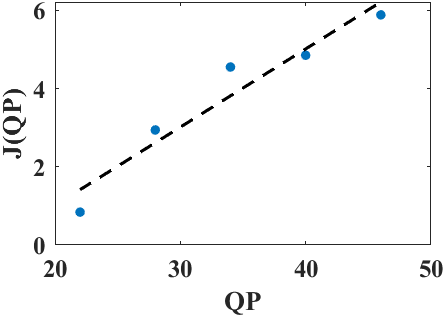}}\hspace{3mm}
    \subfloat[PQS=0.25]{\includegraphics[width=0.2\linewidth]{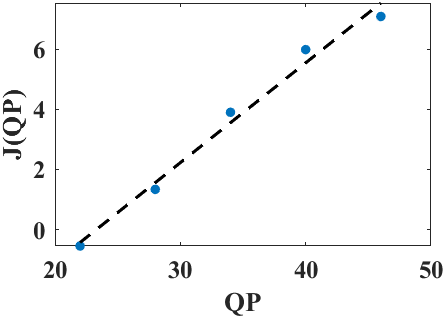}}\hspace{3mm}
    \subfloat[PQS=0.125]{\includegraphics[width=0.2\linewidth]{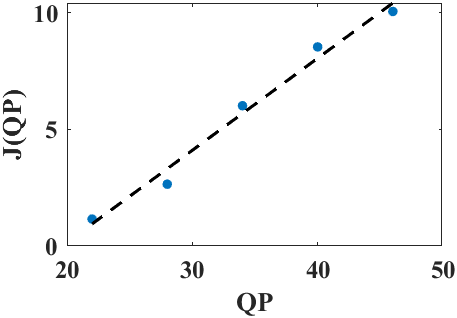}}\hspace{3mm}
        \caption{Relationship between J(QP) and QP at different PQS.}
    \label{fig:J vs TQP}
\end{figure*}

By analyzing Fig.~\ref{fig:MOS_TQS} and Table~\ref{tab:MOSTQSplcc}, we found a significant linear correlation between MOS and TQS. Based on the least squares error criterion, we estimate the predicted MOS by
\begin{equation}
PMOS = \boldsymbol{\alpha} \cdot TQS + \boldsymbol{\beta},
\label{equ:MOS vs TQS}
\end{equation}
where \(\alpha\) is related to the texture distortion quantization parameter, \(\beta\) is related to the geometric distortion quantization parameter and $TQS$ is
\begin{equation}
TQS = 2^\frac{QP-4}{6}.
\label{equ:QP vs TQS}
\end{equation}

\subsection{Estimation of Texture Distortion Factor}
\label{4B}
In Section~\ref{4A}, we deliberated on the influence of texture details on the quality scores of PCs, highlighting that the more abundant the texture details, the more likely higher quality scores can be attained. This phenomenon is depicted in Fig.~\ref{fig:MOS_TQS} as the varying slopes of the relationship curves between MOS and TQS for PCs with varying degrees of texture detail. Therefore, when approximating the texture distortion factor \(\alpha\), it is imperative to account for the interrelation between \(\alpha\) and the parameter indicative of the richness of PC texture details, known as TC. The TC of a PC is reflected by the average standard deviation of pixel values within local blocks of the original PC. We computed the PLCC between \(\alpha\) and TC, and the outcomes are presented in Table~\ref{tab:PLCC of a and TC}.

\begin{table}[t]  
\centering
\caption{PLCC between $\alpha$ and TC.} 
\label{tab:PLCC of a and TC}
\scalebox{0.9}{
\begin{tabular}{c | c c c c}  
\hline
\hline
PQS & 1 & 0.5 & 0.25 & 0.125\\
\hline  
PLCC  & 0.9214 & 0.9082 & 0.7121 & 0.6156\\
\hline
\hline
\end{tabular}}
\end{table}

It was observed that in cases where geometric distortion is not significant, a exhibits a robust linear correlation with TC. Yet, as geometric distortion becomes more pronounced, its influence on the MOS becomes dominant, causing the linear relationship between \(\alpha\) and TC to diminish progressively. We have formulated the relationship between \(\alpha\) and TC with the subsequent function:
\begin{equation}
\boldsymbol{\alpha} = c \cdot TC + d,
\label{equ: B vs TC}
\end{equation}
where coefficients \textit{c} and \textit{d} are derived from a training process utilizing the least squares error criterion.

However, in the context of NR bitstream-layer PCQA models, the original PC is not retrievable, rendering direct acquisition of TC unfeasible. To surmount this challenge, we harness the QP and TBPP extracted from the bitstream information of the compressed PC to estimate TC. Under conditions of constant geometric and textural distortions, the relationship between TC and TBPP is illustrated in Fig.~\ref{fig:TBPP vs. TC}, indicating an approximately linear association. Fig.~\ref{fig:H vs TQP} and Fig.~\ref{fig:J vs TQP} demonstrate that the slope \emph{H} (\ref{equ:TQP vs s}) and intercept \emph{J} (\ref{equ:QP vs i}) of this linear function are dependent on QP. These relationships are encapsulated within the following equations:
\begin{equation}
TC = H(QP) \cdot TBPP + J(QP),
\label{equ:TC vs TBPP}
\end{equation}
\begin{equation}
H(QP) = a_1 \cdot QP^2 + a_2 \cdot QP + a_3,
\label{equ:TQP vs s}
\end{equation}
\begin{equation}
J(QP) = b_1 \cdot QP + b_2,
\label{equ:QP vs i}
\end{equation}
where coefficients $a_{1}$, $a_{2}$, $a_{3}$, $b_{1}$ and $b_{2}$ are derived through a training regimen employing the least squares error criterion.

Thus, the texture distortion factor \(\alpha\) and texture prediction score $PMOS_T$ can be estimated by the following functions:
\begin{equation}
\label{equ:TC}
\begin{split}
\boldsymbol{\alpha} = c \cdot (H(QP) \cdot TBPP + J(QP)) + d
\end{split}
\end{equation}
and
\begin{equation}
PMOS_T(QP,TBPP) = \boldsymbol{\alpha} \cdot TQS,
\label{equ: MOS vs PQS QP}
\end{equation}
\subsection{Estimation of Geometric Distortion Factor and Proposed PCQA Model}\label{4C}
When exploring the relationship between MOS and TQS in Section~\ref{4A}, a situation arose where geometric distortion dominates under conditions of high geometric distortion. This is specifically manifested by the increasing influence of the geometric distortion factor on the model as geometric distortion increases, gradually surpassing the impact of the texture distortion factor on the model. Therefore, it is necessary to establish the relationship between the geometric distortion factor and the quantization parameter PQS, which affects the degree of geometric distortion. 
\begin{table*}[ht]  
\centering
\caption{Parameter Values.} 
\label{tab:Parameter values}
\vspace{-6pt}
\scalebox{0.9}{
\begin{tabular}{c c c c c c c c c c}  
\\
\hline
\hline
\(a_{1}\) & \(a_{2}\) &\(a_{3}\) & \(b_{1}\) & \(b_{2}\) & \(c\) &\(d\) &\(f_1\) &\(f_2\)\\  
\hline  
      0.2176  & -11.1828 & 146.7245 & 0.2428 & -3.2494 & 0.0013 & -0.2042 & -2.7005 & 88.1843\\
\hline
\hline
\end{tabular}  
}
\end{table*}
\begin{figure}\centering
\scalebox{0.8}{
{\includegraphics[width=0.8\linewidth]{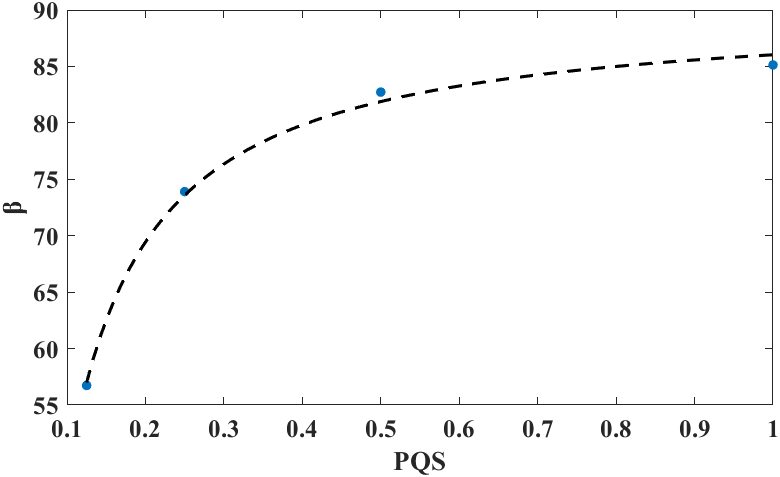}}\hskip0.1em}
\caption{Relationship between $\boldsymbol{\beta}$ and PQS.}
\label{fig:parameters vs PQS}
\end{figure}
\begin{table*}
\centering
\caption{PLCC, SRCC and RMSE performance evaluation of the proposed model against existing models in different Databases. The best performance is shown in \textbf{bold}. The second best is \underline{underlined}.}
\label{tab:performance}
\scalebox{0.69}{
  \begin{tabular}{c |c |c c c c c| c| c c c c}
  \hline
  \hline
  \multirow{2}{*}{Database}&\multirow{2}{*}{Subset}&\multicolumn{5}{c|}{FR}&\multicolumn{1}{c|}{RR}&\multicolumn{4}{c}{NR}\\
  \cline{3-12}
  &&MPED~\cite{yang2022mped} & $\operatorname{PSNR}_{Y}$~\cite{mekuria2016evaluation} & PointSSIM~\cite{alexiou2020towards} & $\operatorname{IW-SSIM}_{P}$~\cite{liu2022perceptual}& GraphSIM~\cite{yang2020inferring} & PCM$_{RR}$~\cite{viola2020reduced} &
   3DTA~\cite{zhu20243dta} & GMS-3DQA~\cite{zhang2024gms} & MM-PCQA~\cite{zhang2022mm} & \textbf{streamPCQ-OR (ours)}\\
   \hline
   \multirow{3}{*}{WPC5.0}
&PLCC$\uparrow$
&0.5858 & 0.7176 & 0.6889 & 0.8310 & 0.8658 & 0.7335 & 0.8622 & \underline{0.9068} & 0.8699 & \textbf{0.9284}\\
&SRCC$\uparrow$
&0.6146 & 0.7321 & 0.6948 & 0.8598 & 0.8806 & 0.7320 & 0.8673 & \underline{0.9070} & 0.8689 & \textbf{0.9324}\\
&RMSE$\downarrow$
&11.1283 & 9.5639 & 9.9537 & 7.6391 & 6.8712 & 13.7314 & 6.9549 & \underline{5.7879} & 6.7736 & \textbf{5.1035}\\
\midrule
\multirow{3}{*}{BASICS~\cite{ak2024basics}}
&PLCC$\uparrow$
&0.8431 & 0.7676 & 0.8170 & 0.8810 & 0.9428 & 0.5363 & 0.8866 & \underline{0.9528} & 0.9214 & \textbf{0.9773}\\
&SRCC$\uparrow$
&0.8150 & 0.7310 & 0.8219 & 0.7459 & 0.9025 & 0.5753 & 0.8419 & \underline{0.9050} & 0.8609 & \textbf{0.9264}\\
&RMSE$\downarrow$
&0.6740 & 0.8034 & 0.7228 & 0.5930 & 0.4179 & 1.2535 & 0.5798 & \underline{0.3807} & 0.4871 & \textbf{0.2656}\\
\midrule
\multirow{3}{*}{M-PCCD~\cite{alexiou2019comprehensive}}
&PLCC$\uparrow$
&0.8065 & 0.7140 & 0.9538 & 0.8190 & \underline{0.9621} & 0.6082 & 0.7366 & \textbf{0.9943} & 0.8889 & 0.9493\\
&SRCC$\uparrow$
&0.8222 & 0.7262 & 0.9514 & 0.7960 & 0.9114 & 0.9250 & 0.6948 & \textbf{0.9744} & 0.8390 & \underline{0.9542}\\
&RMSE$\downarrow$
&0.8540 & 1.0113 & 0.4342 & 0.8288 & \underline{0.3938} & 1.4444 & 0.9769 & \textbf{0.1535} & 0.6617 & 0.4539\\
  \hline
  \hline
  \end{tabular}
  }
\end{table*}

We calculated the average of the intercepts of the MOS versus TQS fitting curves for different PCs under fixed geometric distortion conditions, and then used the following function to estimate the relationship as shown in Fig.~\ref{fig:parameters vs PQS}:
\begin{equation}
\boldsymbol{\beta} = \frac{f_1}{PQS} + f_2,
\label{equ: MOS vs PQS}
\end{equation}
where $f_{1}$ and $f_{2}$ are obtained through training based on the least squares error criterion. The estimated value of $\boldsymbol{\beta}$ can be considered the geometric prediction score $PMOS_G$.

Finally, by integrating the estimated texture and geometric distortion factors with (\ref{equ:MOS vs TQS}), we can derive the proposed streamPCQ-OR model as following:
\begin{equation}
PMOS = c \cdot (H(QP) \cdot TBPP + J(QP)) + d + \frac{f_1}{PQS} + f_2.
\label{equ: MOS vs PQS QP1}
\end{equation}

\section{Experimental Results and Discussion}
\label{5}

\subsection{Parameter Values}
Table~\ref{tab:Parameter values} presents the parameter values. The values of $a_{1}$, $a_{2}$, $a_{3}$, $b_{1}$, and $b_{2}$, were determined using equation (\ref{equ:TQP vs s}) and (\ref{equ:QP vs i}). Parameters $c$ and $d$ were determined based on equation (\ref{equ:TC}), and $f_{1}$ and $f_{2}$ were determined following equation (\ref{equ: MOS vs PQS}). Unless otherwise specified, these parameters remained constant throughout all subsequent experiments reported in this section. Adjustments to these values may be necessary for PCs from other databases.

\subsection{Performance Comparison}
\begin{figure*}
    \centering
    \captionsetup{justification=centering}
    \subfloat[MPED~\cite{yang2022mped}]
    {\includegraphics[width=4.3cm,height=3cm]{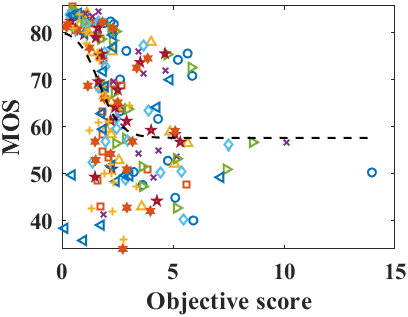}}\hskip0.2em
\subfloat[$\operatorname{PSNR}_{Y}$~\cite{mekuria2016evaluation}]
    {\includegraphics[width=4.3cm,height=3cm]{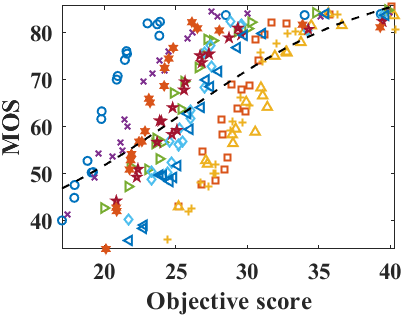}}\hskip0.2em
    \subfloat[PointSSIM~\cite{alexiou2020towards}]
    {\includegraphics[width=4.3cm,height=3cm]{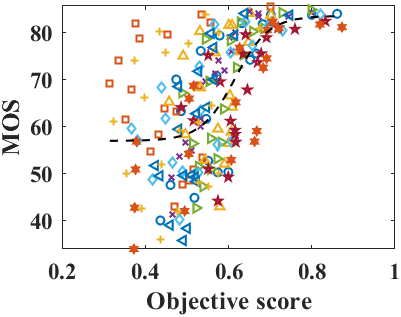}}\hskip0.2em
    \subfloat[$\operatorname{IW-SSIM}_{P}$~\cite{liu2022perceptual}]
    {\includegraphics[width=4.3cm,height=3cm]{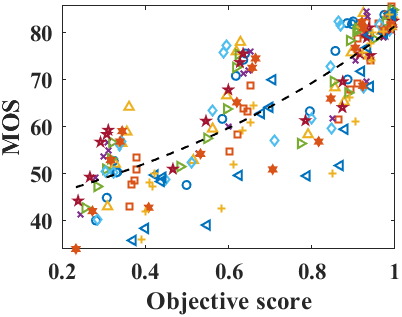}}\hskip0.2em 
    \subfloat[MS-GraphSIM~\cite{zhang2021ms}]
    {\includegraphics[width=4.3cm,height=3cm]{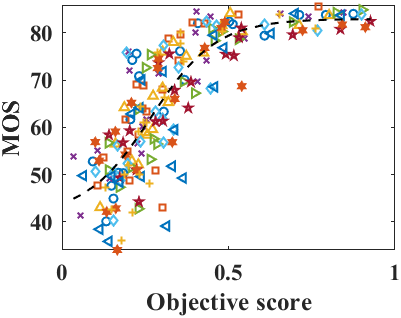}}\hskip0.2em
    \subfloat[GraphSIM~\cite{yang2020inferring}]
    {\includegraphics[width=4.3cm,height=3cm]{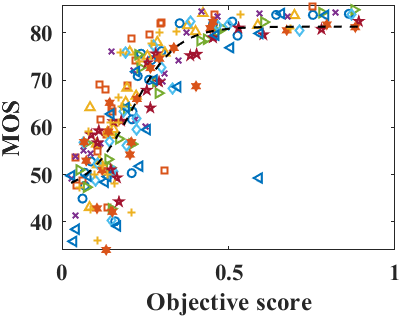}}\hskip0.2em
    \subfloat[$\operatorname{PCM}_{RR}$~\cite{viola2020reduced}]
    {\includegraphics[width=4.3cm,height=3cm]{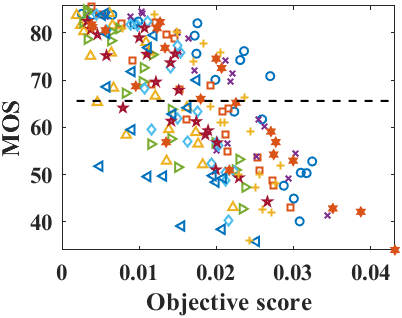}}\hskip0.2em
    \subfloat[3DTA~\cite{zhu20243dta}]
    {\includegraphics[width=4.3cm,height=3cm]{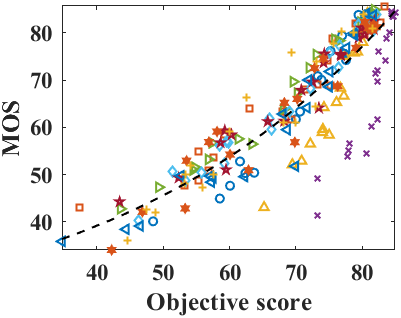}}\hskip0.2em
    \subfloat[GMS-3DQA~\cite{zhang2024gms}]
    {\includegraphics[width=4.3cm,height=3cm]{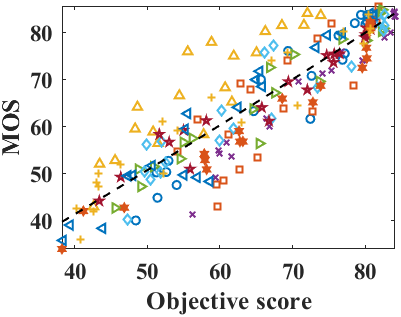}}\hskip0.2em
    \subfloat[MM-PCQA~\cite{zhang2022mm}]
    {\includegraphics[width=4.3cm,height=3cm]{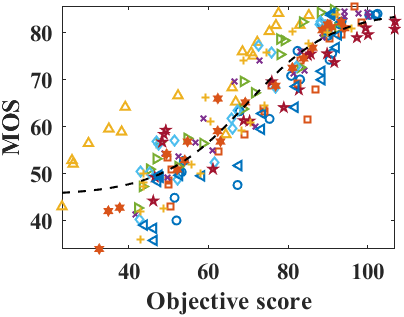}}\hskip0.2em
    \subfloat[\textbf{streamPCQ-OR(ours)}]
    {\includegraphics[width=4.3cm,height=3cm]{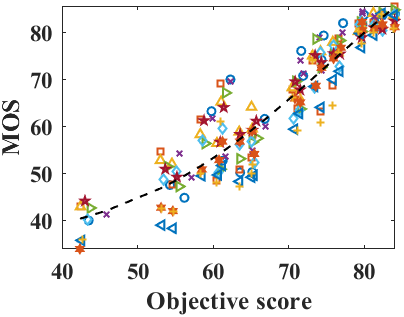}}\hskip0.2em
    {\includegraphics[width=4.3cm,height=3cm]{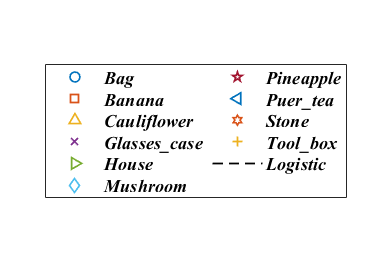}}\hskip0.2em
    \caption{Scatter plots and best fitting logistic functions of objective scores versus MOS on the proposed WPC5.0 database.}
    \label{fig:scatter plots}
\end{figure*}

To evaluate the performance of the model, we selected other competitive PCQA models, namely (A) MPED~\cite{yang2022mped}, (B) $\operatorname{PSNR}_{Y}$~\cite{mekuria2016evaluation}, (C) PointSSIM~\cite{alexiou2020towards}, (D) $\operatorname{IW-SSIM}_{P}$~\cite{liu2022perceptual}, (E) MS-GraphSIM~\cite{zhang2021ms}, (F) GraphSIM~\cite{yang2020inferring}, (G) $\operatorname{PCM}_{RR}$~\cite{viola2020reduced}, (H) 3DTA~\cite{zhu20243dta}, (I) GMS-3DQA~\cite{zhang2024gms} and (J) MM-PCQA~\cite{zhang2022mm} for performance comparison with the proposed streamPCQ-OR model on three databases: WPC5.0, BASICS~\cite{ak2024basics} and M-PCCD~\cite{alexiou2019comprehensive}. Initially, scatter plots were created, as shown in Fig.~\ref{fig:scatter plots}, depicting the objective scores obtained by all selected models on the WPC5.0 database against MOS along the best-fitting logistic function. We compared the performance of the selected PCQA models using three traditional statistical metrics: PLCC, SRCC and root mean square error (RMSE). The results are summarized in Table~\ref{tab:performance}. It can be observed that the streamPCQ-OR demonstrates strong competitiveness compared to other objective models, especially showing excellent predictive performance on the WPC5.0 and BASICS~\cite{ak2024basics} databases. In addition, due to the overlap between the training set SJTU-PCQA~\cite{yang2020predicting} of GMS-3DQA~\cite{zhang2024gms} and the database M-PCCD~\cite{alexiou2019comprehensive}, the performance of streamPCQ-OR on the M-PCCD~\cite{alexiou2019comprehensive} database is slightly lower than that of GMS-3DQA~\cite{zhang2024gms}.

\subsection{Ablation Study}
To investigate the contributions of the model's textural and geometric components to the final model outcomes, an ablation study was conducted. Utilizing the same training and validation procedures as streamPCQ-OR, we developed three models: $PMOS_T (QP)$, $PMOS_T (QP, TBPP)$ and $PMOS_G (PQS)$, the parameters of these models are trained separately. Table~\ref{tab:ablation test} presents the results of the study, highlighting the significant impact of QP, TBPP and PQS on the overall model.
\subsection{Leave-one-out Cross-validation}
In this study, to assess the generalizability of the proposed model and prevent overfitting, we employed a content-sensitive leave-one-out cross-validation (LOOCV) strategy, with results presented in Table~\ref{tab:LOOCV}. Specifically, our dataset comprises 20 distinct reference PCs, denoted as n=20. During the LOOCV process, we select one group of PCs as the validation set for each iteration, while the remaining (n-1) groups serve as the training set. Each validation set consists of 20 distorted PCs, corresponding to a single original PC. We repeat this training and validation procedure n times, with a different group of PCs chosen as the validation set for each iteration, and the model parameters for each iteration are related to the corresponding training. The data in Table~\ref{tab:LOOCV} demonstrate that the proposed model exhibits excellent performance in statistical metrics and possesses strong adaptability and stability.
\begin{table}[t]  
\centering
\caption{Ablation Study on the Proposed WPC5.0 Database.} 
\label{tab:ablation test}
\scalebox{0.9}{\begin{tabular}{c | c c c }  
\hline
\hline
PCQA model & PLCC$\uparrow$ & SRCC$\uparrow$ & RMSE$\downarrow$\\
\hline 
$PMOS_T (QP)$         & 0.5741 & 0.5578 & 11.2429\\
$PMOS_T (QP, TBPP)$   & 0.6177 & 0.6172 & 10.7981\\
$PMOS_G (PQS)$        & 0.7456 & 0.7357 & 9.1511\\
$\textbf{streamPCQ-OR (ours)}$       & \textbf{0.9284} & \textbf{0.9324} & \textbf{5.1035}\\
\hline
\hline
\end{tabular}  
}
\end{table}
\begin{table}[t]  
\centering
\caption{LOOCV Results on the Proposed WPC5.0 Database.} 
\label{tab:LOOCV}
\scalebox{0.9}{
\begin{tabular}{c | c c c }  
\hline
\hline
Name & PLCC$\uparrow$ & SRCC$\uparrow$ & RMSE$\downarrow$\\
\hline
\emph{Bag}             & 0.9581 & 0.9549 & 4.2017 \\
\emph{Banana}          & 0.8955 & 0.8632 & 5.8175 \\
\emph{Biscuits}        & 0.9313 & 0.9398 & 3.8754 \\
\emph{Cake}            & 0.9363 & 0.9233 & 3.4254 \\
\emph{Cauliflower}     & 0.9671 & 0.9654 & 3.0166 \\
\emph{Flowerpot}       & 0.9775 & 0.9549 & 2.6388 \\
\emph{Glasses\_case}   & 0.9687 & 0.9474 & 3.2931 \\
\emph{Honeydew\_melon} & 0.9650 & 0.9594 & 4.5080 \\
\emph{House}           & 0.9709 & 0.9729 & 3.2019 \\
\emph{Litchi}          & 0.9554 & 0.9654 & 4.4973 \\
\emph{Mushroom}        & 0.9656 & 0.9398 & 3.4417 \\
\emph{Pen\_container}  & 0.8958 & 0.9353 & 5.5862 \\
\emph{Pineapple}       & 0.9707 & 0.9624 & 2.7355 \\
\emph{Ping-pong\_bat}  & 0.9817 & 0.9624 & 2.7940 \\
\emph{Puer\_tea}       & 0.9851 & 0.9519 & 2.6285 \\
\emph{Pumpkin}         & 0.9804 & 0.9805 & 2.2438 \\
\emph{Ship}            & 0.9150 & 0.9278 & 5.1377 \\
\emph{Statue}          & 0.9457 & 0.9444 & 4.5923 \\
\emph{Stone}           & 0.9795 & 0.9684 & 2.9313 \\
\emph{Tool\_box}       & 0.9419 & 0.9068 & 4.9602 \\
\hline
Mean                   & 0.9697 & 0.9721 & 4.4540\\
Standard deviation     & 0.0254 & 0.0242 & 1.0231\\
\hline
\hline
\end{tabular}}
\end{table}

\subsection{Robustness Analysis}
\begin{figure}[t]
\centering
{\includegraphics[width=0.48\columnwidth]{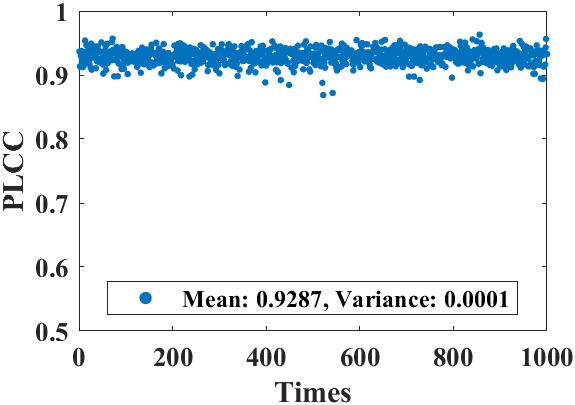}}
  \vspace{1mm}
{\includegraphics[width=0.48\columnwidth]{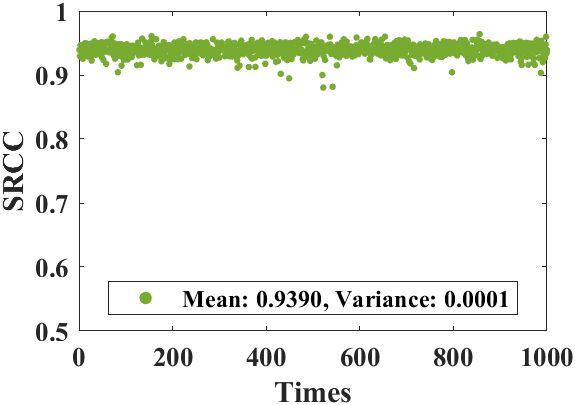}}
\vspace{1mm}
{\includegraphics[width=0.48\columnwidth]{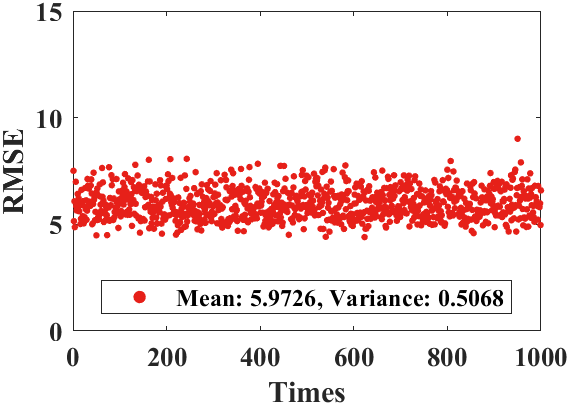}}
  \caption{Robustness of the proposed streamPCQ-OR model on the proposed WPC5.0 database. PLCC, SRCC and RMSE between MOS and objective score are computed for 1000 random splits
of training/validation sets.
}\label{fig:1000plccsrcc}
\end{figure}

To validate the robustness of the proposed streamPCQ-OR model, we randomly selected 10 categories of PCs from the WPC5.0 database as the training set, and the remaining 10 categories as the validation set. Subsequently, we randomly chose 1000 partitioning scenarios from all possible groupings and calculated the PLCC, SRCC and RMSE between the MOS and predictive scores for these 1000 sets of randomly divided training/validation sets. The results depicted in Fig.~\ref{fig:1000plccsrcc} confirm the accuracy of the proposed model.

\subsection{Statistical Significance Analysis}
\begin{figure}[t]
    \centering
    \subfloat[WPC5.0 (proposed)]{\includegraphics[width=0.4\linewidth]{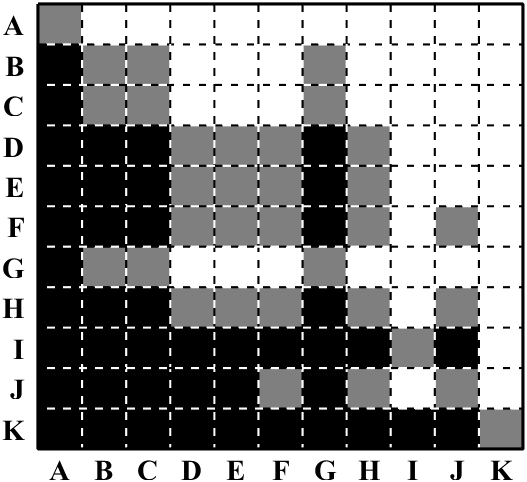}}\hskip1em
    \subfloat[BASICS~\cite{ak2024basics}]{\includegraphics[width=0.4\linewidth]{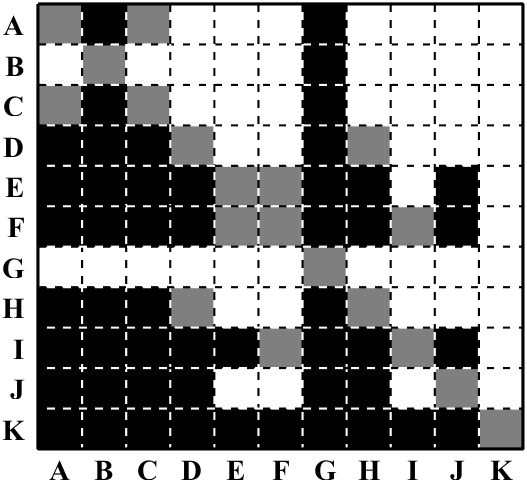}}\hskip1em
    \subfloat[M-PCCD~\cite{alexiou2019comprehensive}]{\includegraphics[width=0.4\linewidth]{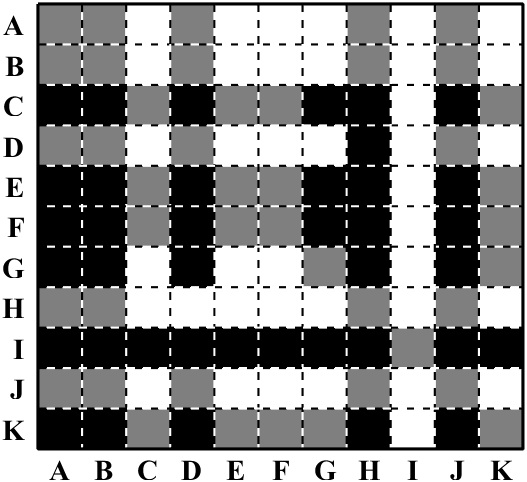}}\hskip1em
    \caption{Statistical significance analysis on three PC databases: (A) $\operatorname{MPED}$~\cite{yang2022mped} (B) $\operatorname{PSNR_Y}$~\cite{mekuria2016evaluation} (C) $\operatorname{PointSSIM}$~\cite{alexiou2020towards} (D) $\operatorname{IW-SSIM_P}$~\cite{liu2022perceptual} (E) $\operatorname{MS-GraphSIM}$~\cite{zhang2021ms} (F) $\operatorname{GraphSIM}$~\cite{yang2020inferring} (G) $\operatorname{PCM_{RR}}$ (H) $\operatorname{3DTA}$~\cite{zhu20243dta} (I) $\operatorname{GMS-3DQA}$~\cite{zhang2024gms} (J) $\operatorname{MM-PCQA}$~\cite{zhang2022mm} (K) \textbf{streamPCQ-OR (ours)}.}
    \label{fig:statistical Significance}
\end{figure}

To verify the statistical significance of the proposed model, we conducted the corresponding analysis following the procedure described in reference~\cite{sheikh2006statistical}. Initially, a nonlinear regression method was applied to map the objective quality scores onto the subjective scores. Throughout this process, the expected value of all prediction errors is zero, hence a model with lower variance is generally considered to be superior to one with higher variance. We performed a hypothesis test using the F-statistic. Given the large sample size, we can reasonably assume that the error terms follow a normal distribution according to the Central Limit Theorem. The test statistic is the ratio of variances. The null hypothesis is that the predictive errors from different quality models are identically distributed and exhibit no statistically significant differences (at a 95\% confidence level). We conducted a comparative analysis for each pair of objective PCQA models and presented the results in Fig.~\ref{fig:statistical Significance}, where a black square indicates that the row model significantly outperforms the column model, a white square indicates the opposite, and a gray square suggests that the row and column models are statistically equivalent in performance. The statistical results indicate that the performance significance of the streamPCQ-OR model on the WPC5.0, BASICS~\cite{ak2024basics} and M-PCCD~\cite{alexiou2019comprehensive} databases is either significantly better or indistinguishable from other objective models.
\subsection{Time Complexity Analysis}
Due to the fact that the bitstream-layer model does not require full decoding, it has a significant advantage in time complexity compared to the media-layer model that requires complete decoding. To make a comparison, we selected three PC samples from the WPC5.0 database, each with different point counts and features, and tested the proposed streamPCQ-OR model for time efficiency against other competitive PCQA models. The results are summarized in Table~\ref{tab:shijianfuza}. These tests were conducted on a high-performance workstation equipped with an Intel® Xeon® Gold 6253CL CPU running at 3.10GHz, 128GB of SAMSUNG MZ7LH480HAHQ-00005 RAM, an ST4000NM000A 2HZ100 hard drive, an NVIDIA GeForce RTX 3090 graphics card, and the Windows 10 Professional operating system. The results presented in Table~\ref{tab:shijianfuza} demonstrate that our model has a significantly lower time complexity compared to other models, which provides strong support for real-time quality monitoring of PC bitstreams during network transmission.
\begin{table}[t]
\centering
\fontsize{6}{7}\selectfont 
\caption{Time Complexity Test Results}
\label{tab:shijianfuza}
\begin{tabular}{
   >{\centering\arraybackslash}p{1.5cm}|
   >{\centering\arraybackslash}m{1.5cm} 
   >{\centering\arraybackslash}p{1.1cm} 
   >{\centering\arraybackslash}p{1.1cm} 
   >{\centering\arraybackslash}p{1.3cm} 
}
\hline
\hline
 Content & PCQA model & Execution time (Seconds) & Execution time ($10^{-6}$ second \ per point) & Execution time \ (Relative to \textbf{streamPCQ-OR}) \\ \hline

\multirow{6}{1.2cm}{\centering Glasses\_case\\(716659 points)} 
&MPED & 171.11 & 238.76& 14044.71  \\
&PSNR$_{Y}$ & 4.92& 6.86& 403.53 
 \\
&PointSSIM & 33.84 & 47.22& 2777.65 
 \\
&IW-SSIM$_{P}$ & 81.25 & 113.38& 6669.41 
 \\
&GraphSIM & 310.70 & 434.54& 25561.16
 \\
&MS-GraphSIM & 13.63 & 19.01& 1118.24 
\\
&PCM$_{RR}$ & 259.09 & 361.53& 21266.47 
 \\ 
&\textbf{streamPCQ-OR} & \textbf{0.012} & \textbf{0.017}
 & \textbf{1} \\  \hline
\multirow{5}{1.4cm}{\centering Honeydew\_melon\\(1431071 points)}  
&MPED & 309.59 & 216.33& 24036.67 
 \\
&PSNR$_{Y}$ & 9.45 & 6.60& 733.33 
 \\
&PointSSIM & 70.06 & 48.96 & 5440 \\
&IW-SSIM$_{P}$ & 121.14 & 84.65 & 9405.55 
 \\
&GraphSIM & 941.59 & 657.96& 73106.67 
 \\
&MS-GraphSIM & 32.38 & 22.62& 2513.33 
 \\
&PCM$_{RR}$ & 522.92 & 365.4& 40600 \\ 
&\textbf{streamPCQ-OR} & \textbf{0.013} & \textbf{0.009} & \textbf{1} \\  \hline
\multirow{5}{1.2cm}{\centering Pen\_container\\(2878318 points)} 
&MPED & 627.86 & 218.13 & 24236.67 
 \\
&PSNR$_{Y}$ & 18.29 & 6.35& 705.56 
\\
&PointSSIM & 143.52 & 49.86& 5540 
\\
&IW-SSIM$_{P}$ & 219.16 & 76.14& 8460 
 \\
&GraphSIM & 4126.87& 1433.78& 159308.89 
 \\
&MS-GraphSIM & 213.84 & 74.29
 & 8254.44 
 \\
&PCM$_{RR}$ & 1305.80 & 453.67& 50407.78 
 \\ 
&\textbf{streamPCQ-OR} & \textbf{0.025} & \textbf{0.009} & \textbf{1} \\  
\hline
\hline
\end{tabular}
\end{table}

\section{Conclusion}\label{6}
In this study, we first constructed the WPC5.0 database with a scale of 400 including 4 geometric multiplied by 5 attitude distortion levels. To the best of our knowledge, it is the first PCQA database dedicated to Octree-RAHT encoding mode. It can be used not only to establish PCQA models specially designed for Octree-RAHT encoding mode but to further study universal PCQA issues. Secondly, we developed the streamPCQ-OR model, the first PCQA model dedicated to Octree-RAHT encoding mode, which is suitable for real-time quality monitoring because of its incomplete decoding nature. After extracting relevant parameters by the bitstream analyzer, considering the masking effect of the human eyes, it sequentially estimates texture complexity, texture distortion factor and geometric distortion factor, and finally obtains the predicted perceptual quality. We compared the proposed streamPCQ-OR model with existing competitive PCQA models and further validated its stability and generalization, finding that it has excellent performance, generalization and robustness. For the convenience of future research, the database and source code will be made publicly available.

\bibliographystyle{IEEEtran}
\bibliography{streamPCQ-OR}

\begin{thebibliography}{10}
\providecommand{\url}[1]{#1}
\csname url@samestyle\endcsname
\providecommand{\newblock}{\relax}
\providecommand{\bibinfo}[2]{#2}
\providecommand{\BIBentrySTDinterwordspacing}{\spaceskip=0pt\relax}
\providecommand{\BIBentryALTinterwordstretchfactor}{4}
\providecommand{\BIBentryALTinterwordspacing}{\spaceskip=\fontdimen2\font plus
\BIBentryALTinterwordstretchfactor\fontdimen3\font minus \fontdimen4\font\relax}
\providecommand{\BIBforeignlanguage}[2]{{%
\expandafter\ifx\csname l@#1\endcsname\relax
\typeout{** WARNING: IEEEtran.bst: No hyphenation pattern has been}%
\typeout{** loaded for the language `#1'. Using the pattern for}%
\typeout{** the default language instead.}%
\else
\language=\csname l@#1\endcsname
\fi
#2}}
\providecommand{\BIBdecl}{\relax}
\BIBdecl

\bibitem{gu20193d}
S.~Gu, J.~Hou, H.~Zeng, H.~Yuan, and K.-K. Ma, ``{3D Point Cloud Attribute Compression Using Geometry-Guided Sparse Representation},'' \emph{IEEE Transactions on Image Processing}, vol.~29, pp. 796--808, 2019.

\bibitem{9234526}
A.~Raake, S.~Borer, S.~M. Satti, J.~Gustafsson, R.~R.~R. Rao, S.~Medagli, P.~List, S.~Göring, D.~Lindero, W.~Robitza, G.~Heikkilä, S.~Broom, C.~Schmidmer, B.~Feiten, U.~Wüstenhagen, T.~Wittmann, M.~Obermann, and R.~Bitto, ``{Multi-Model Standard for Bitstream-, Pixel-Based and Hybrid Video Quality Assessment of UHD/4K: ITU-T P.1204},'' \emph{IEEE Access}, vol.~8, pp. 193\,020--193\,049, 2020.

\bibitem{9194311}
Q.~Liu, H.~Yuan, J.~Hou, R.~Hamzaoui, and H.~Su, ``{Model-Based Joint Bit Allocation Between Geometry and Color for Video-Based 3D Point Cloud Compression},'' \emph{IEEE Transactions on Multimedia}, vol.~23, pp. 3278--3291, 2021.

\bibitem{9106052}
Q.~Liu, H.~Yuan, R.~Hamzaoui, and H.~Su, ``{Coarse to fine rate control for region-based 3D point cloud compression},'' in \emph{2020 IEEE International Conference on Multimedia \& Expo Workshops (ICMEW)}.\hskip 1em plus 0.5em minus 0.4em\relax IEEE, 2020, pp. 1--6.

\bibitem{liu2019comprehensive}
H.~Liu, H.~Yuan, Q.~Liu, J.~Hou, and J.~Liu, ``{A comprehensive study and comparison of core technologies for MPEG 3-D point cloud compression},'' \emph{IEEE Transactions on Broadcasting}, vol.~66, no.~3, pp. 701--717, 2019.

\bibitem{ISOIEC2019GPCC}
{MPEG 3DG}, ``{Geometry-Based Point Cloud Compression},'' in \emph{{ISO/IEC JTC 1/SC29/WG11 Doc. N18474, Geneva, Switzerland,}}, 2019.

\bibitem{ISOIEC2018VPCC}
{MPEG 3DG}, ``{Video-Based Point Cloud Compression},'' in \emph{{ISO/IEC JTC 1/SC 29/WG 11 Doc. N18030, Macau, China,}}, 2018.

\bibitem{whiteG-PCC}
{Ohji Nakagami and Sebastien Lasserre and Sugio Toshiyasu and Marius Preda}, ``{White paper on G-PCC},'' in \emph{{ISO/IEC JTC 1/SC 29/AG 03 Doc. N0111, Antalya,}}, 2023.

\bibitem{ak2024basics}
A.~Ak, E.~Zerman, M.~Quach, A.~Chetouani, A.~Smolic, G.~Valenzise, and P.~Le~Callet, ``{BASICS: Broad Quality Assessment of Static Point Clouds in a Compression Scenario},'' \emph{IEEE Transactions on Multimedia}, 2024.

\bibitem{alexiou2019comprehensive}
E.~Alexiou, I.~Viola, T.~M. Borges, T.~A. Fonseca, R.~L. De~Queiroz, and T.~Ebrahimi, ``{A comprehensive study of the rate-distortion performance in MPEG point cloud compression},'' \emph{APSIPA Transactions on Signal and Information Processing}, vol.~8, p. e27, 2019.

\bibitem{mekuria2016evaluation}
R.~Mekuria, Z.~Li, C.~Tulvan, and P.~Chou, ``{Evaluation Criteria for PCC (Point Cloud Compression)},'' \emph{ISO/IEC JTC}, vol.~1, p. N16332, 2016.

\bibitem{tian2017geometric}
D.~Tian, H.~Ochimizu, C.~Feng, R.~Cohen, and A.~Vetro, ``{Geometric distortion metrics for point cloud compression},'' in \emph{2017 IEEE International Conference on Image Processing (ICIP)}.\hskip 1em plus 0.5em minus 0.4em\relax IEEE, 2017, pp. 3460--3464.

\bibitem{mekuria2017performance}
R.~Mekuria, S.~Laserre, and C.~Tulvan, ``{Performance assessment of point cloud compression},'' in \emph{2017 IEEE Visual Communications and Image Processing (VCIP)}.\hskip 1em plus 0.5em minus 0.4em\relax IEEE, 2017, pp. 1--4.

\bibitem{alexiou2020towards}
E.~Alexiou and T.~Ebrahimi, ``{Towards a Point Cloud Structural Similarity Metric},'' in \emph{2020 IEEE International Conference on Multimedia \& Expo Workshops (ICMEW)}.\hskip 1em plus 0.5em minus 0.4em\relax IEEE, 2020, pp. 1--6.

\bibitem{liu2022perceptual}
Q.~Liu, H.~Su, Z.~Duanmu, W.~Liu, and Z.~Wang, ``{Perceptual Quality Assessment of Colored 3D Point Clouds},'' \emph{IEEE Transactions on Visualization and Computer Graphics}, vol.~29, no.~8, pp. 3642--3655, 2022.

\bibitem{yang2020inferring}
Q.~Yang, Z.~Ma, Y.~Xu, Z.~Li, and J.~Sun, ``{Inferring Point Cloud Quality via Graph Similarity},'' \emph{IEEE transactions on pattern analysis and machine intelligence}, vol.~44, no.~6, pp. 3015--3029, 2020.

\bibitem{zhang2021ms}
Y.~Zhang, Q.~Yang, and Y.~Xu, ``{MS-GraphSIM: Inferring Point Cloud Quality via Multiscale Graph Similarity},'' in \emph{Proceedings of the 29th ACM International Conference on Multimedia}, 2021, pp. 1230--1238.

\bibitem{meynet2020pcqm}
G.~Meynet, Y.~Nehm{\'e}, J.~Digne, and G.~Lavou{\'e}, ``{PCQM: A Full-Reference Quality Metric for Colored 3D Point Clouds},'' in \emph{2020 Twelfth International Conference on Quality of Multimedia Experience (QoMEX)}.\hskip 1em plus 0.5em minus 0.4em\relax IEEE, 2020, pp. 1--6.

\bibitem{meynet2019pc}
G.~Meynet, J.~Digne, and G.~Lavou{\'e}, ``{PC-MSDM: A quality metric for 3D point clouds},'' in \emph{2019 Eleventh International Conference on Quality of Multimedia Experience (QoMEX)}.\hskip 1em plus 0.5em minus 0.4em\relax IEEE, 2019, pp. 1--3.

\bibitem{viola2020color}
I.~Viola, S.~Subramanyam, and P.~Cesar, ``{A Color-Based Objective Quality Metric for Point Cloud Contents},'' in \emph{2020 Twelfth International Conference on Quality of Multimedia Experience (QoMEX)}.\hskip 1em plus 0.5em minus 0.4em\relax IEEE, 2020, pp. 1--6.

\bibitem{hua2020vqa}
L.~Hua, M.~Yu, G.~Jiang, Z.~He, and Y.~Lin, ``{VQA-CPC: A novel visual quality assessment metric of color point clouds},'' in \emph{Optoelectronic Imaging and Multimedia Technology VII}, vol. 11550.\hskip 1em plus 0.5em minus 0.4em\relax SPIE, 2020, pp. 244--252.

\bibitem{hua2022cpc}
L.~Hua, M.~Yu, Z.~He, R.~Tu, and G.~Jiang, ``{CPC-GSCT: Visual quality assessment for coloured point cloud based on geometric segmentation and colour transformation},'' \emph{IET Image Processing}, vol.~16, no.~4, pp. 1083--1095, 2022.

\bibitem{wang2024integrated}
Y.~Wang, X.~Yao, P.~Zhu, W.~Li, M.~Cao, and Q.~Hu, ``Integrated heterogeneous graph attention network for incomplete multi-modal clustering,'' \emph{International Journal of Computer Vision}, pp. 1--20, 2024.

\bibitem{huang2022class}
H.~Huang, Y.~Wang, Q.~Hu, and M.-M. Cheng, ``Class-specific semantic reconstruction for open set recognition,'' \emph{IEEE transactions on pattern analysis and machine intelligence}, vol.~45, no.~4, pp. 4214--4228, 2022.

\bibitem{javaheri2020improving}
A.~Javaheri, C.~Brites, F.~Pereira, and J.~Ascenso, ``{Improving Psnr-Based Quality Metrics Performance For Point Cloud Geometry},'' in \emph{2020 IEEE International Conference on Image Processing (ICIP)}.\hskip 1em plus 0.5em minus 0.4em\relax IEEE, 2020, pp. 3438--3442.

\bibitem{javaheri2020generalized}
A.~Javaheri, C.~Brites, F.~Pereira, and J.~Ascenso, ``{A Generalized Hausdorff Distance Based Quality Metric for Point Cloud Geometry},'' in \emph{2020 Twelfth International Conference on Quality of Multimedia Experience (QoMEX)}.\hskip 1em plus 0.5em minus 0.4em\relax IEEE, 2020, pp. 1--6.

\bibitem{diniz2020multi}
R.~Diniz, P.~G. Freitas, and M.~C. Farias, ``{Multi-Distance Point Cloud Quality Assessment},'' in \emph{2020 IEEE International Conference on Image Processing (ICIP)}.\hskip 1em plus 0.5em minus 0.4em\relax IEEE, 2020, pp. 3443--3447.

\bibitem{diniz2020towards}
R.~Diniz, P.~G. Freitas, and M.~C. Farias, ``{Towards a Point Cloud Quality Assessment Model using Local Binary Patterns},'' in \emph{2020 Twelfth International Conference on Quality of Multimedia Experience (QoMEX)}.\hskip 1em plus 0.5em minus 0.4em\relax IEEE, 2020, pp. 1--6.

\bibitem{diniz2021novel}
R.~Diniz, P.~G. Freitas, and M.~Farias, ``{A novel point cloud quality assessment metric based on perceptual color distance patterns},'' \emph{Electronic Imaging}, vol.~33, pp. 1--11, 2021.

\bibitem{diniz2020local}
R.~Diniz, P.~G. Freitas, and M.~C. Farias, ``{Local Luminance Patterns for Point Cloud Quality Assessment},'' in \emph{2020 IEEE 22nd International Workshop on Multimedia Signal Processing (MMSP)}.\hskip 1em plus 0.5em minus 0.4em\relax IEEE, 2020, pp. 1--6.

\bibitem{diniz2021color}
R.~Diniz, P.~G. Freitas, and M.~C. Farias, ``{Color and Geometry Texture Descriptors for Point-Cloud Quality Assessment},'' \emph{IEEE Signal Processing Letters}, vol.~28, pp. 1150--1154, 2021.

\bibitem{xu2021epes}
Y.~Xu, Q.~Yang, L.~Yang, and J.-N. Hwang, ``{EPES: Point Cloud Quality Modeling Using Elastic Potential Energy Similarity},'' \emph{IEEE Transactions on Broadcasting}, vol.~68, no.~1, pp. 33--42, 2021.

\bibitem{torlig2018novel}
E.~M. Torlig, E.~Alexiou, T.~A. Fonseca, R.~L. de~Queiroz, and T.~Ebrahimi, ``A novel methodology for quality assessment of voxelized point clouds,'' in \emph{Applications of Digital Image Processing XLI}, vol. 10752.\hskip 1em plus 0.5em minus 0.4em\relax SPIE, 2018, pp. 174--190.

\bibitem{wu2021subjective}
X.~Wu, Y.~Zhang, C.~Fan, J.~Hou, and S.~Kwong, ``{Subjective Quality Database and Objective Study of Compressed Point Clouds With 6DoF Head-Mounted Display},'' \emph{IEEE Transactions on Circuits and Systems for Video Technology}, vol.~31, no.~12, pp. 4630--4644, 2021.

\bibitem{he2021towards}
Z.~He, G.~Jiang, Z.~Jiang, and M.~Yu, ``{Towards A Colored Point Cloud Quality Assessment Method Using Colored Texture And Curvature Projection},'' in \emph{2021 IEEE International Conference on Image Processing (ICIP)}.\hskip 1em plus 0.5em minus 0.4em\relax IEEE, 2021, pp. 1444--1448.

\bibitem{yang2020predicting}
Q.~Yang, H.~Chen, Z.~Ma, Y.~Xu, R.~Tang, and J.~Sun, ``{Predicting the Perceptual Quality of Point Cloud: A 3D-to-2D Projection-Based Exploration},'' \emph{IEEE Transactions on Multimedia}, vol.~23, pp. 3877--3891, 2020.

\bibitem{he2022tgp}
Z.~He, G.~Jiang, M.~Yu, Z.~Jiang, Z.~Peng, and F.~Chen, ``{TGP-PCQA: Texture and geometry projection based quality assessment for colored point clouds},'' \emph{Journal of Visual Communication and Image Representation}, vol.~83, p. 103449, 2022.

\bibitem{tu2023pseudo}
R.~Tu, G.~Jiang, M.~Yu, Y.~Zhang, T.~Luo, and Z.~Zhu, ``{Pseudo-Reference Point Cloud Quality Measurement Based on Joint 2-D and 3-D Distortion Description},'' \emph{IEEE Transactions on Instrumentation and Measurement}, 2023.

\bibitem{viola2020reduced}
I.~Viola and P.~Cesar, ``{A Reduced Reference Metric for Visual Quality Evaluation of Point Cloud Contents},'' \emph{IEEE Signal Processing Letters}, vol.~27, pp. 1660--1664, 2020.

\bibitem{liu2021reduced}
Q.~Liu, H.~Yuan, R.~Hamzaoui, H.~Su, J.~Hou, and H.~Yang, ``{Reduced Reference Perceptual Quality Model With Application to Rate Control for Video-Based Point Cloud Compression},'' \emph{IEEE Transactions on Image Processing}, vol.~30, pp. 6623--6636, 2021.

\bibitem{liu2022reduced}
Y.~Liu, Q.~Yang, and Y.~Xu, ``{Reduced Reference Quality Assessment for Point Cloud Compression},'' in \emph{2022 IEEE International Conference on Visual Communications and Image Processing (VCIP)}.\hskip 1em plus 0.5em minus 0.4em\relax IEEE, 2022, pp. 1--5.

\bibitem{10375131}
H.~Su, Q.~Liu, H.~Yuan, Q.~Cheng, and R.~Hamzaoui, ``{Support Vector Regression-Based Reduced- Reference Perceptual Quality Model for Compressed Point Clouds},'' \emph{IEEE Transactions on Multimedia}, vol.~26, pp. 6238--6249, 2024.

\bibitem{zhou2023reduced}
W.~Zhou, G.~Yue, R.~Zhang, Y.~Qin, and H.~Liu, ``{Reduced-Reference Quality Assessment of Point Clouds via Content-Oriented Saliency Projection},'' \emph{IEEE Signal Processing Letters}, vol.~30, pp. 354--358, 2023.

\bibitem{zhang2024reduced}
Z.~Zhang, Y.~Zhou, C.~Li, K.~Fu, W.~Sun, X.~Liu, X.~Min, and G.~Zhai, ``{A Reduced-Reference Quality Assessment Metric for Textured Mesh Digital Humans},'' in \emph{ICASSP 2024-2024 IEEE International Conference on Acoustics, Speech and Signal Processing (ICASSP)}.\hskip 1em plus 0.5em minus 0.4em\relax IEEE, 2024, pp. 2965--2969.

\bibitem{zhang2022no}
Z.~Zhang, W.~Sun, X.~Min, T.~Wang, W.~Lu, and G.~Zhai, ``{No-Reference Quality Assessment for 3D Colored Point Cloud and Mesh Models},'' \emph{IEEE Transactions on Circuits and Systems for Video Technology}, vol.~32, no.~11, pp. 7618--7631, 2022.

\bibitem{liu2023point}
Y.~Liu, Q.~Yang, Y.~Xu, and L.~Yang, ``{Point Cloud Quality Assessment: Dataset Construction and Learning-based No-reference Metric},'' \emph{ACM Transactions on Multimedia Computing, Communications and Applications}, vol.~19, no.~2s, pp. 1--26, 2023.

\bibitem{liu2022progressive}
Q.~Liu, Y.~Liu, H.~Su, H.~Yuan, and R.~Hamzaoui, ``{Progressive Knowledge Transfer Based on Human Visual Perception Mechanism for Perceptual Quality Assessment of Point Clouds},'' \emph{arXiv preprint arXiv:2211.16646}, 2022.

\bibitem{zhou2024blind}
W.~Zhou, Q.~Yang, W.~Chen, Q.~Jiang, G.~Zhai, and W.~Lin, ``{Blind Quality Assessment of Dense 3D Point Clouds with Structure Guided Resampling},'' \emph{ACM Transactions on Multimedia Computing, Communications and Applications}, 2024.

\bibitem{zhu20243dta}
L.~Zhu, J.~Cheng, X.~Wang, H.~Su, H.~Yang, H.~Yuan, and J.~Korhonen, ``{3DTA: No-Reference 3D Point Cloud Quality Assessment with Twin Attention},'' \emph{IEEE Transactions on Multimedia}, 2024.

\bibitem{Shan_2024_CVPR}
Z.~Shan, Y.~Zhang, Q.~Yang, H.~Yang, Y.~Xu, J.-N. Hwang, X.~Xu, and S.~Liu, ``{Contrastive Pre-Training with Multi-View Fusion for No-Reference Point Cloud Quality Assessment},'' in \emph{Proceedings of the IEEE/CVF Conference on Computer Vision and Pattern Recognition (CVPR)}, June 2024, pp. 25\,942--25\,951.

\bibitem{tu2022v}
R.~Tu, G.~Jiang, M.~Yu, T.~Luo, Z.~Peng, and F.~Chen, ``{V-PCC Projection Based Blind Point Cloud Quality Assessment for Compression Distortion},'' \emph{IEEE Transactions on Emerging Topics in Computational Intelligence}, vol.~7, no.~2, pp. 462--473, 2022.

\bibitem{shan2023gpa}
Z.~Shan, Q.~Yang, R.~Ye, Y.~Zhang, Y.~Xu, X.~Xu, and S.~Liu, ``{GPA-Net:No-Reference Point Cloud Quality Assessment With Multi-Task Graph Convolutional Network},'' \emph{IEEE Transactions on Visualization and Computer Graphics}, 2023.

\bibitem{liu2021pqa}
Q.~Liu, H.~Yuan, H.~Su, H.~Liu, Y.~Wang, H.~Yang, and J.~Hou, ``{PQA-Net: Deep No Reference Point Cloud Quality Assessment via Multi-View Projection},'' \emph{IEEE transactions on circuits and systems for video technology}, vol.~31, no.~12, pp. 4645--4660, 2021.

\bibitem{zhang2024gms}
Z.~Zhang, W.~Sun, H.~Wu, Y.~Zhou, C.~Li, Z.~Chen, X.~Min, G.~Zhai, and W.~Lin, ``{GMS-3DQA: Projection-Based Grid Mini-patch Sampling for 3D Model Quality Assessment},'' \emph{ACM Transactions on Multimedia Computing, Communications and Applications}, vol.~20, no.~6, pp. 1--19, 2024.

\bibitem{liu2023once}
\BIBentryALTinterwordspacing
Y.~Liu, Q.~Yang, Y.~Zhang, Y.~Xu, L.~Yang, X.~Xu, and S.~Liu, ``{Once-Training-All-Fine: No-Reference Point Cloud Quality Assessment via Domain-relevance Degradation Description},'' 2023. [Online]. Available: \url{https://arxiv.org/abs/2307.01567}
\BIBentrySTDinterwordspacing

\bibitem{tao2021point}
W.-x. Tao, G.-y. Jiang, Z.-d. Jiang, and M.~Yu, ``{Point Cloud Projection and Multi-Scale Feature Fusion Network Based Blind Quality Assessment for Colored Point Clouds},'' in \emph{Proceedings of the 29th ACM International Conference on Multimedia}, 2021, pp. 5266--5272.

\bibitem{yang2022no}
Q.~Yang, Y.~Liu, S.~Chen, Y.~Xu, and J.~Sun, ``{No-Reference Point Cloud Quality Assessment via Domain Adaptation},'' in \emph{Proceedings of the IEEE/CVF Conference on Computer Vision and Pattern Recognition (CVPR)}, June 2022, pp. 21\,179--21\,188.

\bibitem{fan2022no}
Y.~Fan, Z.~Zhang, W.~Sun, X.~Min, N.~Liu, Q.~Zhou, J.~He, Q.~Wang, and G.~Zhai, ``{A No-reference Quality Assessment Metric for Point Cloud Based on Captured Video Sequences},'' in \emph{2022 IEEE 24th International Workshop on Multimedia Signal Processing (MMSP)}.\hskip 1em plus 0.5em minus 0.4em\relax IEEE, 2022, pp. 1--5.

\bibitem{zhang2022treating}
Z.~{Zhang}, W.~{Sun}, Y.~{Zhu}, X.~{Min}, W.~{Wu}, Y.~{Chen}, and G.~{Zhai}, ``{Evaluating Point Cloud from Moving Camera Videos: A No-Reference Metric},'' \emph{arXiv e-prints}, p. arXiv:2208.14085, Aug. 2022.

\bibitem{wang2024zoom}
J.~Wang, W.~Gao, and G.~Li, ``{Zoom to Perceive Better: No-Reference Point Cloud Quality Assessment via Exploring Effective Multiscale Feature},'' \emph{IEEE Transactions on Circuits and Systems for Video Technology}, 2024.

\bibitem{hua2021bqe}
L.~Hua, G.~Jiang, M.~Yu, and Z.~He, ``{BQE-CVP: Blind Quality Evaluator for Colored Point Cloud Based on Visual Perception},'' in \emph{2021 IEEE International Symposium on Broadband Multimedia Systems and Broadcasting (BMSB)}.\hskip 1em plus 0.5em minus 0.4em\relax IEEE, 2021, pp. 1--6.

\bibitem{zhang2022mm}
\BIBentryALTinterwordspacing
Z.~Zhang, W.~Sun, X.~Min, Q.~Zhou, J.~He, Q.~Wang, and G.~Zhai, ``{MM-PCQA: Multi-Modal Learning for No-reference Point Cloud Quality Assessment},'' 2023. [Online]. Available: \url{https://arxiv.org/abs/2209.00244}
\BIBentrySTDinterwordspacing

\bibitem{chen2024dynamic}
W.~Chen, Q.~Jiang, W.~Zhou, L.~Xu, and W.~Lin, ``{Dynamic Hypergraph Convolutional Network for No-Reference Point Cloud Quality Assessment},'' \emph{IEEE Transactions on Circuits and Systems for Video Technology}, 2024.

\bibitem{chai2024plain}
X.~Chai, F.~Shao, B.~Mu, H.~Chen, Q.~Jiang, and Y.-S. Ho, ``{Plain-PCQA: No-Reference Point Cloud Quality Assessment by Analysis of Plain Visual and Geometrical Components},'' \emph{IEEE Transactions on Circuits and Systems for Video Technology}, 2024.

\bibitem{MU2024122953}
\BIBentryALTinterwordspacing
B.~Mu, F.~Shao, H.~Chen, Q.~Jiang, L.~Xu, and Y.-S. Ho, ``{Hallucinated-PQA: No reference point cloud quality assessment via injecting pseudo-reference features},'' \emph{Expert Systems with Applications}, vol. 243, p. 122953, 2024. [Online]. Available: \url{https://www.sciencedirect.com/science/article/pii/S0957417423034553}
\BIBentrySTDinterwordspacing

\bibitem{liu2022no}
Q.~Liu, H.~Su, T.~Chen, H.~Yuan, and R.~Hamzaoui, ``No-reference bitstream-layer model for perceptual quality assessment of v-pcc encoded point clouds,'' \emph{IEEE Transactions on Multimedia}, vol.~25, pp. 4533--4546, 2022.

\bibitem{su2023bitstream}
H.~Su, Q.~Liu, Y.~Liu, H.~Yuan, H.~Yang, Z.~Pan, and Z.~Wang, ``{Bitstream-Based Perceptual Quality Assessment of Compressed 3D Point Clouds},'' \emph{IEEE Transactions on Image Processing}, vol.~32, pp. 1815--1828, 2023.

\bibitem{10637704}
J.~Lv, H.~Su, Q.~Liu, and H.~Yuan, ``{No-reference Bitstream-based Perceptual Quality Assessment of Octree-Lifting Encoded 3D Point Clouds},'' \emph{IEEE Transactions on Visualization and Computer Graphics}, pp. 1--15, 2024.

\bibitem{alexiou2020pointxr}
E.~Alexiou, N.~Yang, and T.~Ebrahimi, ``{PointXR: A Toolbox for Visualization and Subjective Evaluation of Point Clouds in Virtual Reality},'' in \emph{2020 Twelfth International Conference on Quality of Multimedia Experience (QoMEX)}.\hskip 1em plus 0.5em minus 0.4em\relax IEEE, 2020, pp. 1--6.

\bibitem{series2012methodology}
B.~Series, ``{Methodology for the subjective assessment of the quality of television pictures},'' \emph{Recommendation ITU-R BT}, vol. 500, no.~13, 2012.

\bibitem{YanaComparison}
\BIBentryALTinterwordspacing
Y.~Nehm\'{e}, J.-P. Farrugia, F.~Dupont, P.~LeCallet, and G.~Lavou\'{e}, ``{Comparison of subjective methods, with and without explicit reference, for quality assessment of 3D graphics},'' in \emph{ACM Symposium on Applied Perception 2019}, ser. SAP '19.\hskip 1em plus 0.5em minus 0.4em\relax New York, NY, USA: Association for Computing Machinery, 2019. [Online]. Available: \url{https://doi.org/10.1145/3343036.3352493}
\BIBentrySTDinterwordspacing

\bibitem{van1995quality}
A.~M. Van~Dijk, J.-B. Martens, and A.~B. Watson, ``{Quality asessment of coded images using numerical category scaling},'' in \emph{Advanced Image and Video Communications and Storage Technologies}, vol. 2451.\hskip 1em plus 0.5em minus 0.4em\relax SPIE, 1995, pp. 90--101.

\bibitem{yang2022mped}
Q.~Yang, Y.~Zhang, S.~Chen, Y.~Xu, J.~Sun, and Z.~Ma, ``{MPED: Quantifying Point Cloud Distortion Based on Multiscale Potential Energy Discrepancy},'' \emph{IEEE Transactions on Pattern Analysis and Machine Intelligence}, vol.~45, no.~5, pp. 6037--6054, 2022.

\bibitem{sheikh2006statistical}
H.~R. Sheikh, M.~F. Sabir, and A.~C. Bovik, ``{A Statistical Evaluation of Recent Full Reference Image Quality Assessment Algorithms},'' \emph{IEEE Transactions on image processing}, vol.~15, no.~11, pp. 3440--3451, 2006.

\end{thebibliography}
\end{document}